\documentclass{elsart}
\usepackage{graphicx,amsmath,amssymb}
\usepackage{hyperref}

\def\tr{\qopname\relax{no}{Tr}}

\def\dnabla{\stackrel{\leftrightarrow}{\nabla}}

    \def\<{\langle}         \def\>{\rangle}
      
    \def\tr{{\rm tr}}

        \def\del{\delta}        
        
       \def\down{\downarrow}

  \def\V0{{\mathbf 0}}

 \def\Bj{{\mathbf j}}

\def\Bx{{\mathbf x}}  \def\B0{{\mathbf 0}}

\def\be{\begin{equation}}       \def\ee{\end{equation}}
\def\bea{\begin{eqnarray}}      \def\eea{\end{eqnarray}}

\begin{document}

\begin{frontmatter}



\title{Unconventional interaction between vortices \\
in a polarized Fermi gas }


\author{Vladimir M. Stojanovi\'c}
\address{Department of Physics, Carnegie Mellon University,
Pittsburgh, PA $15213$, USA}
\author{W. Vincent Liu}
\address{Department of Physics and Astronomy, University of
Pittsburgh, Pittsburgh, PA 15260, USA} \corauth[cor]{Corresponding
author.} \ead{wvliu@pitt.edu}
\author{Yong Baek Kim}
\address{Department of Physics, University of Toronto, Toronto, Ontario
M5S 1A7, Canada} \address{Department of Physics, University of
California, Berkeley, CA 94720, USA}

\begin{abstract}

Recently, a homogeneous superfluid state with a single gapless Fermi
surface was predicted to be the ground state of an ultracold Fermi gas
with spin population imbalance in the regime of molecular
Bose-Einstein condensation.  We study vortices in this novel state
using a symmetry-based effective field theory, which captures the
low-energy physics of gapless fermions and superfluid phase
fluctuations. This theory is applicable to all spin-imbalanced
ultracold Fermi gases in the superfluid regime, regardless of whether
the original fermion pairing interaction is weak or strong. We find a
remarkable, unconventional form of the interaction between vortices.
The presence of gapless fermions gives rise to a spatially oscillating
potential, akin to the RKKY indirect-exchange interaction in
non-magnetic metals. We compare the parameters of the effective theory
to the experimentally measurable quantities and further discuss the
conditions for the verification of the predicted new feature. Our
study opens up an interesting question as to the nature of the vortex
lattice resulting from the competition between the usual repulsive
logarithmic (2D Coulomb) and predominantly attractive fermion-induced
interactions.
\end{abstract}

\end{frontmatter}

\section{Introduction}

The appearance of quantized vortices is a hallmark of superfluid
flow. Vortices have been studied for decades, experimentally
and/or theoretically, in a variety of systems as diverse as
type-II superconductors, superfluid Helium liquids, rotating
ultracold atomic Bose and Fermi gases, and neutron
stars.~\cite{vortexbook} Among these systems, the quantum gas of
resonantly interacting fermionic atoms with equal populations of
both (hyperfine) spin components, a prototype system for the
interesting BEC-BCS crossover physics,
~\cite{Leggett:BEC-BCS:80,Nozieres-Schmitt-Rink:85,sademeloranderia:93,engelbrecht:97,Ohashi+Griffin-BCS-BECcrossover:02}
has been intensively studied over the past several years. The
first experimental observation of vortices in 2005 by the MIT
group provided a definitive evidence for superfluidity in atomic
Fermi gases.~\cite{Zwierlein+Ketterle:05} In addition, several
theoretical studies have analyzed the possible new properties of
vortices across a Feshbach resonance from the BCS to the BEC
side.~\cite{bulgac+yu:03,Machida:05,sensarma+ho:06,Botelho:06}

The physics of atomic Fermi gases is also of fundamental interest
beyond the standard BCS/BEC physics, owing to the new tuning
flexibility in the atomic gas systems. Under the condition of
density imbalance (hence mismatched Fermi surfaces) between the
spin-up and -down fermions, a modulated
Larkin-Ovchinnikov-Fulde-Ferrell (LOFF)~\cite{LOFF:65+64}
superfluid phase, has long been theoretically anticipated. The
theoretical interest in pairing with mismatched Fermi surfaces has
been revitalized by the proposal of breached-pairing
superfluidity, with a number of exotic superfluid states being
proposed or revisited.~\cite{Wu-Yip:03,Bedaque:03prl}
Breached-pairing superfluid phase with two gapless Fermi surfaces
(BP2), related to the unstable Sarma phase,~\cite{Sarma:63} was
found to be stable under the introduction of new effects, such as
the mass imbalance and/or momentum-dependent pairing
interaction.~\cite{Liu-Wilczek:03,Forbes+:05} Important
developments in the subject are recent studies by various
groups~\cite{Pao+Wu+Yip:06,Sheehy:06,Son-Stephanov:06}
investigating the Feshbach-resonant regime of strong interactions.
~\cite{Sheehy+Radzihovsky:07} The first experiments on ultracold
fermionic gases with spin population imbalance have recently been
carried out~\cite{ketterle:06,hulet:06} and thereby brought the
subject to the forefront of the cold atom physics. The imbalanced
Fermi gas is presently the subject of fervent research
activities.~\cite{kunyang:05pre,Yi-Duan:06,Iskin:06,Gubankova:06,Parish:07}

One of the states commonly found in various theoretical approaches
~\cite{Pao+Wu+Yip:06,Sheehy:06,Son-Stephanov:06} is a homogeneous
superfluid with a single gapless Fermi surface on the molecular
(BEC) side of the Feshbach resonance (the BEC regime). This state,
which consists of coexisting molecular superfluid and
fully-polarized Fermi gas of the majority-spin component, is
closely related to the BP2 phase,~\cite{Liu-Wilczek:03,Forbes+:05}
but differs from the latter in the number of gapless Fermi
surfaces. We will refer to this phase as BP1 (breached-pairing
state with a single gapless Fermi surface) after
Ref.~\cite{Yi-Duan:06} and \cite{Parish:07} (also dubbed as
``magnetized superfluid'' and denoted SF$_{\textrm{M}}$ in
Ref.~\cite{Sheehy:06} and \cite{Sheehy+Radzihovsky:07}). The BP1
phase is predicted to robustly exist in a relatively large area in
the phase diagram of the spin imbalanced Fermi gases (also called
``polarized Fermi gases''). Additionally, several theoretical
works have found the analog of BP1 in a trap, the so-called
superfluid-normal-mixture phase.~\cite{TrapMixedPhases:06} It is
then of great interest to examine properties of this novel
superfluid.

When fermionic excitations are fully gapped, the physics of
vortices belongs to the universality class of the XY model where
the phase of the superfluid order parameter plays the dominant
role. The vortex sector of this model is described by a $2$D
Coulomb gas of ``charges'' with a repulsive logarithmic
interaction. This paradigm is challenged in a fundamental fashion
in the BP1 state. The presence of gapless fermionic quasiparticles
is the distinguishing feature of this superfluid phase and is
expected to have important consequences for its physical
properties. Since the observation of a vortex lattice is perhaps
the only unambiguous signature of superfluidity in ultracold
fermionic gases,~\cite{Zwierlein+Ketterle:05} it is of interest to
examine possible ramifications of the presence of gapless fermions
on the interaction between vortices in this system.

In this work, we determine the effective interaction potential
between the vortices in the BP1 phase. Because we are solely
concerned with the intrinsic effect of the gapless fermions on the
interaction between vortices in the BP1 state, we consider only
the homogeneous case, disregarding the effects of traps. We
exploit the method of effective field theory,~\cite{weinbergbook}
based explicitly on broken continuous
symmetry.~\cite{schakel:06pre} This method is particularly
suitable for problems involving strongly-coupled systems in the
long-wavelength limit and has already proven to be fruitful in
treatments of strongly-interacting regimes of ultracold atomic
gases.~\cite{andersen:04,liu:eft:06,sonwingate:06,nishida+son:06}
In the present context, the relevant degrees of freedom for a
low-energy effective field theory are the superfluid phase field
and the field describing the gapless fermionic excitations.

We show that the resulting interaction between vortices in the BP1
phase is not of the pure Coulomb form, but contains an additional
fermion-induced contribution that oscillates on a length scale set
by the spin polarization, closely resembling the
Ruderman-Kittel-Kasuya-Yosida (RKKY) indirect-exchange interaction
in non-magnetic metals.~\cite{RKKY}
In order to show that such an unusual vortex interaction
is perfectly compatible with the BP1 phase, we calculate
the superfluid density from a microscopic model in the parameter
regime relevant for the BP1 state. We demonstrate that the superfluid
density in BP1 is positive throughout, which corroborates the
dynamical stability of this phase and warrants its further
investigation.

The outline of the remainder of this paper is as follows. In
Sec.~II we introduce notation and conventions for fermion
quasiparticles and superfluid phase to be used throughout. In Sec.
II we present self-contained field-theoretical derivation of the
superfluid density, followed by the calculation of this quantity
in the parameter regime where the BP1 phase is realized. Sec.~IV
starts with a low-energy effective field theory for gapless
fermions and superfluid phase field, from which we derive an
effective theory for phase fluctuations by integrating out the
fermionic degrees of freedom. In Sec.~V we first derive the
effective theory for vortices and their effective interaction in
momentum space. Then we present the calculation of the effective
vortex interaction potential in real space, accompanied by the
discussion of its physical significance. Finally, we summarize in
Sec.~VI. Some mathematical details are relegated to the
Appendices.

\section{Notation and conventions}
\subsection{Gapless branch of fermionic quasiparticles}\label{fn}

The Bogoliubov-quasiparticle energy spectrum of the system
containing two fermion gases with equal masses ($m$) and unequal
chemical potentials ($\mu_{\uparrow}\neq\mu_{\downarrow}$) is
given by (in what follows $\hbar=1$, unless stated otherwise)
\begin{equation}
E_{\mathbf k}^{\pm}=\sqrt{(\frac{\mathbf{k}^2}{2m}
-\mu)^2+\Delta^{2}}\pm \frac{\delta}{2} \:\:,
\end{equation}
\noindent where $\mu=(\mu_{\uparrow}+\mu_{\downarrow})/2$ is the
average chemical potential (thermodynamically conjugate to the
overall atomic number of two species combined) and
$\delta=\mu_{\downarrow}-\mu_{\uparrow}$ measures the mismatch
between the two chemical potentials (conjugate to the relative
density imbalance). Since our treatment concerns the
spin-polarized homogeneous superfluid realized deeply on the BEC
side of the Feshbach resonance, the average chemical potential is
assumed to be negative ($\mu<0$) in what follows. For
definiteness, we hereafter also assume that $\delta>0$.

\begin{figure}
\begin{center}
\includegraphics[width=0.9\linewidth]{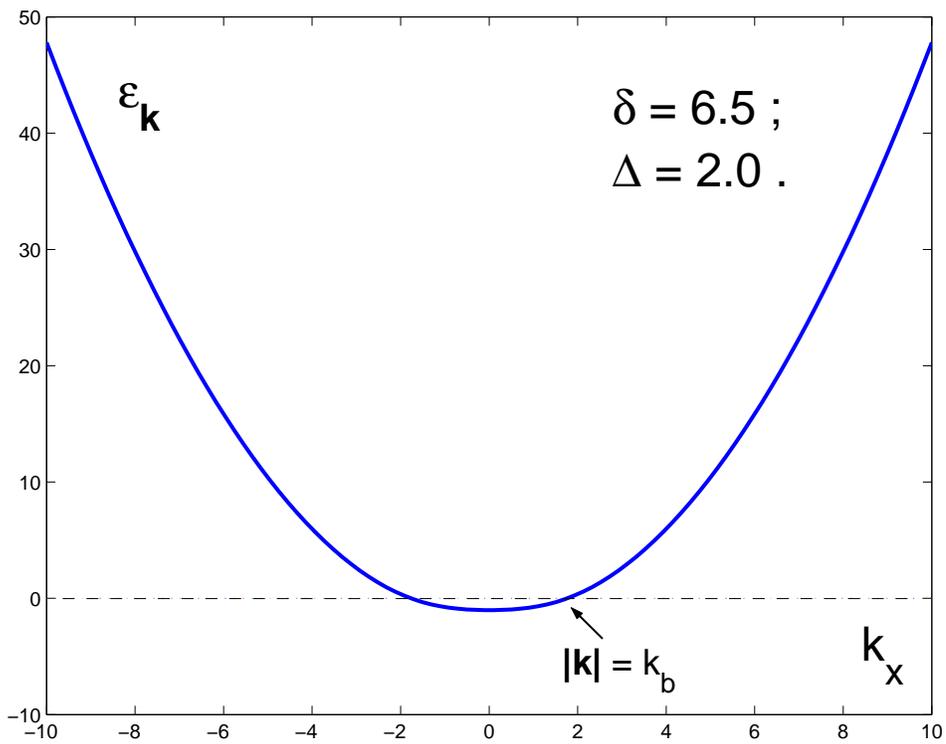}
\end{center}
\caption{An example of gapless fermion quasiparticle dispersion
$\varepsilon_{\mathbf{k}}$. Values of parameters $\delta$ and
$\Delta$ are indicated (expressed in units of
$|\mu|$).}\label{fig:bogoliub}
\end{figure}

For sufficiently large mismatch $\delta$ (more precisely, for
$\delta/2>\sqrt{\mu^2+\Delta^{2}})$, the lower branch $E_{\mathbf
k}^{-}$ of the above quasiparticle dispersion is gapless, with a
single effective Fermi surface. Hereafter, for the sake of
brevity, we denote it as
\begin{equation}\label{disp}
\varepsilon_{\mathbf k}=\sqrt{(\frac{\mathbf{k}^2}{2m}
-\mu)^2+\Delta^{2}}-\frac{\delta}{2} \:\:.
\end{equation}
\noindent The effective Fermi surface is defined by
$\varepsilon_{\mathbf k}=0$ for $|\mathbf{k}|=k_{b}$, where
$k_{b}$ is the radius of the ``breached-pairing Fermi ball'' in
momentum space. The latter is controlled by the density imbalance
$n_{b}=n_{\downarrow}-n_{\uparrow}$ between the two pseudo-spin
components (here $n_{\downarrow}>n_{\uparrow}$, as a consequence
of the fact that $\delta>0$), which,  as implied by the Luttinger
theorem~\cite{Sachdev+Yang:06}, is equal to the volume of the
``breached-pairing Fermi ball'' in momentum space
\begin{equation}
 \int_{|{\mathbf k}|\in [0,k_{b}]}\frac{d^{3}{\mathbf k}}{(2\pi)^{3}}
  =\frac{k_{b}^{3}}{\:\:6\pi^{2}}\:\:.
\end{equation}
\noindent This leads to a simple expression for $k_{b}$ :
\begin{equation}\label{kb}
k_{b}=[6\pi^{2}(n_{\downarrow}-n_{\uparrow})]^{1/3}\:\:.
\end{equation}
\noindent An example of gapless dispersion given by Eq.
\eqref{disp} is depicted in Fig.~\ref{fig:bogoliub}.

\subsection{Superfluid phase field and its decomposition}

The superfluid phase field $\theta(\mathbf{x},\tau)$
represents the phase of the complex Cooper-pair amplitude :
\begin{equation}\label{phasedef}
\langle\psi_{\uparrow}\psi_{\downarrow}\rangle =
|\langle\psi_{\uparrow}\psi_{\downarrow}\rangle|\:e^{i\theta}
\:\:,
\end{equation}
\noindent where $\psi_{\uparrow}$ and $\psi_{\downarrow}$ are the
fields describing fermions of opposite spins. In the following, we use
the standard decomposition of the superfluid phase
field into a static (classical) contribution
$\theta_{v}(\mathbf{x})$ (singular part) and a quantum-fluctuating
contribution $\phi(\mathbf{x},\tau)$ (regular part):
\begin{equation}
\theta(\mathbf{x},\tau)=\theta_{v}(\mathbf{x})+\phi(\mathbf{x},
\tau)\:\:.
\end{equation}
\noindent $\phi(\mathbf{x},\tau)$ is a non-compact (unbounded)
field describing ``spin-wave'' (smooth, i.e., non-topological)
phase fluctuations ; $\theta_{v}(\mathbf{x})$ is a multi-valued
field pertaining to the topological defects of broken global
$U(1)$ symmetry---vortices.

Superfluid vortices can effectively be treated as classical
point-objects in two dimensions. In reality, a two-dimensional
theory is applicable to rotating superfluids as long as the
rotation frequency is not too high and the flow is everywhere
confined to a plane perpendicular to the rotation
axis.~\cite{pismenbook} These conditions allow formation of
straight vortex lines parallel to the axis of rotation. The
arrangement is then essentially two-dimensional, equivalent to an
array of point vortices with circulation of the same sign.

The gradient of the singular part of the phase field can
conveniently be expressed as~\cite{nelsonbook}
\begin{equation}\label{sigrad}
\nabla \theta_{v}=\kappa_{0}(\hat{\mathbf{e}}_{z}\times
\nabla)\int d^{2}\mathbf{x}'\:G(\mathbf{x,x'})
\rho(\mathbf{x}')\:\:,
\end{equation}
\noindent where $\kappa_{0}=h/(2m)=\pi\hbar/m$ ($\hbar$ restored
for the sake of clarity) is the circulation quantum,
$\rho(\mathbf{x})$ stands for the vortex ``charge density'', and
$G(\mathbf{x,x'})=G(|\mathbf{x}-\mathbf{x'}|)$ is the Green's
function of the two-dimensional Laplacian:
\begin{equation}
 \nabla^{2}G(\mathbf{x,x'})=\delta^{(2)}(\mathbf{x}-\mathbf{x'})\:\:.
\end{equation}
The vortex charge density is defined as
\begin{equation}
 \rho(\mathbf{x})=2\pi\sum_{\alpha}N_{\alpha}\delta^{(2)}(\mathbf{x}-
 \mathbf{x}_{\alpha})\:\:,
\end{equation}
where $N_{\alpha}$ is the winding number, viz. topological charge,
of a vortex located at position $\mathbf{x}_{\alpha}$ in the
xy-plane and  $\hat{\mathbf{e}}_{z}$ is the unit vector
perpendicular to this plane. Following the standard
prescription,~\cite{schakel:06pre,kleinert:defect} the gradient of
$\theta_{v}$ can be associated to a vortex gauge (vector) field
$\mathbf{a}(\mathbf{x})$ through $\mathbf{a}=-\nabla \theta_{v}$,
which by virtue of Eq. \eqref{sigrad} becomes
\begin{equation}
\mathbf{a}=-\kappa_{0}(\hat{\mathbf{e}}_{z}\times\nabla)\int
d^{2}\mathbf{x}'\:G(\mathbf{x,x'})\rho(\mathbf{x}') \:\:.
\end{equation}
\noindent Consequently, the vortex gauge field obeys the condition
\begin{equation}\label{jedna}
\nabla \times \mathbf{a}=-\kappa_{0}
\rho(\mathbf{x})\hat{\mathbf{e}}_{z} \:\:,
\end{equation}
\noindent whose momentum-space version (obtained by spatial
Fourier transformation) reads
\begin{equation}\label{dru}
 -i\mathbf{q}\times \mathbf{a}_{\mathbf{q}}=
 \kappa_{0}\tilde{\rho}(\mathbf{q})\hat{\mathbf{e}}_{z} \:\:,
\end{equation}
\noindent where $\mathbf{q}$ is a two-dimensional wave-vector
($\mathbf{q}\cdot\hat{\mathbf{e}}_{z}=0$) of phase fluctuations
and $\tilde{\rho}(\mathbf{q})$ is the Fourier transform of the
vortex charge density $\rho(\mathbf{x})$.

We now make use of the decomposition
$\mathbf{a}_{\mathbf{q}}=\mathbf{a}_{\mathbf{q}}^{\|}
+\mathbf{a}_{\mathbf{q}}^{\bot}$ of the vortex gauge field into
the longitudinal and transverse components with respect to the
momentum $\mathbf{q}$, respectively, and they are given by
\begin{eqnarray}
 \mathbf{a}_{\mathbf{q}}^{\|}=\frac{(\mathbf{q}
 \cdot\mathbf{a}_{\mathbf{q}})
 \mathbf{q}}{\mathbf{q}^{2}}\:\:\:\:,
\qquad \mathbf{a}_{\mathbf{q}}^{\bot}= \frac{(\mathbf{q}\times
\mathbf{a}_{\mathbf{q}})\times \mathbf{q}}{\mathbf{q}^{2}}
\:\:\:\:.
\end{eqnarray}
\noindent Because $\mathbf{q}\times
\mathbf{a}_{\mathbf{q}}^{\|}=0$, Eq. \eqref{dru} implies that
\begin{eqnarray}\label{cond}
\tilde{\rho}(\mathbf{q})\hat{\mathbf{e}}_{z}=
-\frac{i}{\kappa_{0}}\:\mathbf{q}\times
\mathbf{a}_{\mathbf{q}}^{\bot}\:\:, \qquad
\tilde{\rho}(\mathbf{-q})\hat{\mathbf{e}}_{z}=
\frac{i}{\kappa_{0}}\:\mathbf{q}\times
\mathbf{a}_{\mathbf{-q}}^{\bot} \:\:,
\end{eqnarray}
\noindent where the second equation in \eqref{cond} has been
obtained from the first one by a simple replacement
$\mathbf{q}\rightarrow -\mathbf{q}$. The scalar product of the
last two equations yields
\begin{equation}
\tilde{\rho}(\mathbf{q})
\tilde{\rho}(\mathbf{-q})=\frac{1}{\kappa_{0}^{2}}\:
(\mathbf{q}\times \mathbf{a}_{\mathbf{q}}^{\bot})
\cdot(\mathbf{q}\times \mathbf{a}_{\mathbf{-q}}^{\bot})\:\:.
\end{equation}
\noindent Now using the fact that $\mathbf{q}\cdot
\mathbf{a}_{\mathbf{q}}^{\bot}= \mathbf{q}\cdot
\mathbf{a}_{\mathbf{-q}}^{\bot}=0$, together with the identity of
vector algebra
\begin{equation}
(\mathbf{A}\times\mathbf{B})\cdot(\mathbf{C}\times\mathbf{D})
=(\mathbf{A}\cdot\mathbf{C})(\mathbf{B}\cdot\mathbf{D})-
(\mathbf{A}\cdot\mathbf{D})(\mathbf{B}\cdot\mathbf{C})\:\:,
\end{equation}
\noindent it is straightforward to obtain a useful relation
\begin{equation}\label{korisna}
 \tilde{\rho}(\mathbf{q})\tilde{\rho}(\mathbf{-q})=
 \frac{\mathbf{q}^{2}}
 {\kappa_{0}^{2}}\:(\mathbf{a}_{\mathbf{q}}^{\bot}
 \cdot \mathbf{a}_{\mathbf{-q}}^{\bot})\:\:.
\end{equation}

For convenience, we henceforth adopt the Coulomb gauge
$\nabla\cdot \mathbf{a}=0$. In momentum space this reads
$\mathbf{q}\cdot \mathbf{a}_{\mathbf{q}}=0$, which means that in
this gauge the vector field $\mathbf{a}_{\mathbf{q}}$ is purely
transverse ($\mathbf{a}_{\mathbf{q}}^{\|}=0$, viz.
$\mathbf{a}_{\mathbf{q}}=\mathbf{a}_{\mathbf{q}}^{\bot}$).
Consequently, Eq. \eqref{korisna} can be rewritten as
\begin{equation}\label{ind}
\tilde{\rho}(\mathbf{q})\tilde{\rho}(\mathbf{-q})=
\frac{\mathbf{q}^{2}}
{\kappa_{0}^{2}}\:(\mathbf{a}_{\mathbf{q}}\cdot
\mathbf{a}_{\mathbf{-q}})\:\:,
\end{equation}
\noindent a form that will be used in the following sections.

\section{Superfluid density calculation} \label{sdcalc}

The superfluid density is a quantity of paramount importance in
the realm of quantum liquids. This macroscopic observable has the
nature of a transport coefficient and describes the response of a
superfluid system to a Galilean boost transformation. Its
low-temperature behavior reflects the key properties of the ground
state.~\cite{Iida-Baym:01} In what follows, we calculate the
superfluid density in the BP1 state, to be subsequently used as an
input to the effective field theory in the second part of this
paper. In order to make the presentation self-contained, we start
from a microscopic fermion-pairing model and derive a general
expression for the superfluid density by following the standard
field-theoretic method of Ref.~\cite{hejin:06}. This approach has
proved to yield equivalent results as that of
Ref.~\cite{Wu-Yip:03}. We then specialize to the case of equal
mass fermions and evaluate the superfluid density in the relevant
parameter regime for the realization of the BP1 phase.

Our starting point is the microscopic Lagrangian
\begin{equation}\label{micrlagr}
\mathcal{L}_{0}=\sum_{\sigma=\uparrow,\downarrow}\psi_{\sigma}^{*}
\left(\partial_{\tau}-\frac{\nabla^{2}}{2m_{\sigma}}
-\mu_{\sigma}\right)\psi_{\sigma}+g\psi_{\uparrow}^{*}
\psi_{\downarrow}^{*}\psi_{\downarrow}\psi_{\uparrow}
\end{equation}
\noindent describing pairing of two species (denoted by a formal
pseudo-spin variable $\sigma=\uparrow,\downarrow$) of fermions
with masses $m_{\sigma}$, chemical potentials $\mu_{\sigma}$, and
attractive inter-species contact interaction with coupling
constant $g$. In the mean-field approximation, the thermodynamic
potential for this system is given by
\begin{equation}
 \Omega=- \frac{\Delta^{2}}{g}-\beta^{-1}
 \sum_{i\omega_{n}}\int\frac{d^{3}{\mathbf k}}{(2\pi)^{3}}
 \:\tr\ln\mathcal{G}^{-1} \:\:,
\end{equation}
\noindent where
\begin{equation}
\mathcal{G}^{-1}=\left(\begin{array}{rr}
i\omega_{n}-\epsilon_{{\mathbf k},\uparrow} &
\Delta\:\:\:\:\:\:\:\:\:\:\: \\
\Delta\:\:\:\:\:\:\:\:\:\: & i\omega_{n}+
\epsilon_{{\mathbf k},\downarrow} \\
\end{array}\right) \:\:
\end{equation}
\noindent is the inverse of the fermion propagator in the Nambu
space, with free fermion dispersions $\epsilon_{{\mathbf
k},\sigma}={\mathbf k}^{2}/(2m_{\sigma})-\mu_{\sigma}$ and
$\Delta$ set to be real. This fermion propagator has the matrix
form
\begin{equation}\label{propagator}
\mathcal{G}=\left(\begin{array}{rr}
\mathcal{G}_{\uparrow\uparrow} & \mathcal{G}_{\uparrow\downarrow}\\
\mathcal{G}_{\downarrow\uparrow} & \mathcal{G}_{\downarrow\downarrow} \\
\end{array}\right) \:\:
\end{equation}
\noindent with
\begin{eqnarray}
\mathcal{G}_{\uparrow\uparrow}&=&
\frac{i\omega_{n}-\epsilon_{\mathbf{k}}^{-}+
\epsilon_{\mathbf{k}}^{+}}{(i\omega_{n}-
\epsilon_{\mathbf{k}}^{-})^{2}-\epsilon_{\Delta}^{2}} \:\:,\\
\mathcal{G}_{\downarrow\downarrow} &=&
\frac{i\omega_{n}-\epsilon_{\mathbf{k}}^{-}-
\epsilon_{\mathbf{k}}^{+}}{(i\omega_{n}-
\epsilon_{\mathbf{k}}^{-})^{2}-\epsilon_{\Delta}^{2}} \:\:,\\
\mathcal{G}_{\uparrow\downarrow}   &=&
-\frac{\Delta}{(i\omega_{n}-
\epsilon_{\mathbf{k}}^{-})^{2}-\epsilon_{\Delta}^{2}} \:\:,\\
\mathcal{G}_{\downarrow\uparrow}   &=&
-\frac{\Delta}{(i\omega_{n}-
\epsilon_{\mathbf{k}}^{-})^{2}-\epsilon_{\Delta}^{2}} \:\:,
\end{eqnarray}
\noindent and $\epsilon_{\mathbf{k}}^{\pm}$, $\epsilon_{\Delta}$
defined as
\begin{equation}\label{}
\epsilon_{\mathbf{k}}^{\pm}=\frac{1}{2}\left(\epsilon_{{\mathbf
k},\uparrow} \pm \epsilon_{{\mathbf k},\downarrow}\right)\qquad,
\qquad\epsilon_{\Delta}=\sqrt{\epsilon_{\mathbf{k}}^{+2}+\Delta^{2}}
\:\:.
\end{equation}
\noindent Quasiparticle energy spectrum is determined by the poles
of propagator \eqref{propagator}, i.e., by the solution of
equation det $\mathcal{G}^{-1}=0$ :
\begin{equation}\label{spectrum}
E_{\mathbf{k}}^{+}=\epsilon_{\Delta}+
\epsilon_{\mathbf{k}}^{-}\qquad, \qquad
E_{\mathbf{k}}^{-}=\epsilon_{\Delta}-
\epsilon_{\mathbf{k}}^{-}\:\:.
\end{equation}

The occupation numbers of two species of fermions can be
calculated from the diagonal elements of propagator
\eqref{propagator} :
\begin{eqnarray}\label{}
n_{\uparrow}(\mathbf{k}) &=& \beta^{-1} \lim_{\eta\rightarrow
0}\sum_{i\omega_{n}}
\mathcal{G}_{\uparrow\uparrow}e^{i\omega_{n}\eta} \:\:,\\
n_{\downarrow}(\mathbf{k}) &=& -\beta^{-1}\lim_{\eta\rightarrow
0}\sum_{i\omega_{n}}\mathcal{G}_{\downarrow\downarrow}
e^{-i\omega_{n}\eta}\:\:.
\end{eqnarray}
\noindent Upon performing Matsubara frequency summations we obtain
\begin{eqnarray}
n_{\uparrow}(\mathbf{k})   &=&
u_{\mathbf{k}}^{2}\:n_{F}(E^{+}_{\mathbf k})+
v_{\mathbf{k}}^{2}\:n_{F}(-E^{-}_{\mathbf k}) \:\:, \\
n_{\downarrow}(\mathbf{k}) &=&
u_{\mathbf{k}}^{2}\:n_{F}(E^{-}_{\mathbf k})+
v_{\mathbf{k}}^{2}\:n_{F}(-E^{+}_{\mathbf k})\:\:,
\end{eqnarray}
\noindent where $n_{F}(z)\equiv (\exp(\beta z)+1)^{-1}$ is the
Fermi distribution function and
\begin{equation}\label{}
u_{\mathbf{k}}^{2}=\frac{1}{2}
\left(1+\frac{\epsilon_{\mathbf{k}}^{+}}{\epsilon_{\Delta}}\right)
\qquad, \qquad v_{\mathbf{k}}^{2}=\frac{1}{2}
\left(1-\frac{\epsilon_{\mathbf{k}}^{+}}{\epsilon_{\Delta}}\right)
\:\:,
\end{equation}
\noindent are the coherence factors (squared Bogoliubov
amplitudes).

Under Galilean boost with velocity $\mathbf{v}_{s}$, $\Delta$
transforms as $\Delta\rightarrow\Delta e^{i(m_{\downarrow}+
m_{\uparrow})\mathbf{v}_{s}\cdot\mathbf{x}}$, while fermion fields
transform as $\psi_{\sigma}\rightarrow\psi_{\sigma}
e^{i\mathbf{q}_{\sigma}\cdot\mathbf{x}}$, where
$\mathbf{q}_{\sigma}=m_{\sigma}\mathbf{v}_{s}$. The superfluid
(mass) density tensor $\rho_{ij}$ is defined through
\begin{equation}
\Omega(\mathbf{v}_{s})=\Omega(0)
+\mathbf{j}_{s}\cdot\mathbf{v}_{s}
+\frac{1}{2}\rho_{ij}(\mathbf{v}_{s})_{i}
(\mathbf{v}_{s})_{j}+\mathcal{O}(\mathbf{v}_{s}^{3}) \:\:.
\end{equation}
\noindent For a homogeneous and isotropic superfluid this tensor
is diagonal, viz. $\rho_{ij}=\delta_{ij}\rho_{s}/3$, where
$\rho_{s}$ is the superfluid mass density. Accordingly, the last
formula reduces to
\begin{equation}
\Omega(\mathbf{v}_{s})=\Omega(0)
+\mathbf{j}_{s}\cdot\mathbf{v}_{s}
+\frac{1}{6}\rho_{s}\mathbf{v}_{s}^{2}
+\mathcal{O}(\mathbf{v}_{s}^{3}) \:\:.
\end{equation}

By transforming the fermion fields and $\Delta$ according to the
rules stated above, the thermodynamic potential becomes
\begin{equation}\label{therpot}
\Omega(\mathbf{v}_{s})=-\frac{\Delta^{2}}{g}-\beta^{-1}
\sum_{i\omega_{n}}\int\frac{d^{3}{\mathbf k}}{(2\pi)^{3}}
\:\tr\ln\mathcal{G}_{s}^{-1} \:\:,
\end{equation}
\noindent where $\mathcal{G}_{s}^{-1}(i\omega_{n},\mathbf{k})$ is
the $\mathbf{v}_{s}$-dependent fermion inverse propagator
\begin{equation}\label{}
\mathcal{G}_{s}^{-1}=\left(\begin{array}{rr}
i\omega_{n}-\epsilon_{\mathbf{k}+m_{\uparrow}
\mathbf{v}_{s},\uparrow} & \Delta\qquad \\
\Delta\qquad & i\omega_{n}+\epsilon_{\mathbf{k}
-m_{\downarrow}\mathbf{v}_{s},\downarrow}\\
\end{array}\right) \:\:.
\end{equation}
\noindent It is easy to check that the latter can be expressed as
\begin{equation}\label{}
\mathcal{G}_{s}^{-1}=\mathcal{G}^{-1}-(\mathbf{k}\cdot
\mathbf{v}_{s})\mathbf{1}_{2\times2}-
\frac{1}{2}\mathbf{v}_{s}^{2}\Sigma_{m}\:\:,
\end{equation}
\noindent where
$\Sigma_{m}=\textrm{diag}(m_{\uparrow},-m_{\downarrow})$, i.e., as
\begin{equation}\label{}
\mathcal{G}_{s}^{-1}=\mathcal{G}^{-1}\left\{1-(\mathbf{k}\cdot
\mathbf{v}_{s})\mathcal{G}-\frac{1}{2}
\mathbf{v}_{s}^{2}(\mathcal{G}\Sigma_{m})\right\}\:\:.
\end{equation}
\noindent By making use of the well-known expansion formula
$\ln(1-z)=z+\frac{1}{2}z^{2}+\mathcal{O}(z^{3})$, we obtain
\begin{equation}\label{}
\tr\ln\mathcal{G}_{s}^{-1}=\tr\ln\mathcal{G}^{-1}+
(\mathbf{k}\cdot\mathbf{v}_{s})\tr(\mathcal{G})-
\frac{\mathbf{v}_{s}^{2}}{2}\tr(\mathcal{G}\Sigma_{m})
-\frac{1}{2}(\mathbf{k}\cdot\mathbf{v}_{s})^{2}
\tr(\mathcal{G}^{2})+\mathcal{O}(\mathbf{v}_{s}^{3}) \:\:.
\end{equation}
\noindent The last result, combined with Eq. \eqref{therpot},
enables us to expand the thermodynamic potential
$\Omega(\mathbf{v}_{s})$ in powers of $\mathbf{v}_{s}$ and read
off the superfluid density from the quadratic term :
\begin{equation}\label{}
\rho_{s}=m_{\downarrow}n_{\downarrow}+
m_{\uparrow}n_{\uparrow}+\int\frac{d^{3}{\mathbf k}}
{(2\pi)^{3}}\frac{{\mathbf k}^{2}}{3}(\sigma_{\uparrow\uparrow}
+\sigma_{\downarrow\downarrow}+2\sigma_{\uparrow\downarrow})\:\:,
\end{equation}
\noindent where
\begin{eqnarray}
\sigma_{\uparrow\uparrow} &=& \beta^{-1}\sum_{i\omega_{n}}
\mathcal{G}_{\uparrow\uparrow}\mathcal{G}_{\uparrow\uparrow}\:\:,\\
\sigma_{\downarrow\downarrow} &=& \beta^{-1}\sum_{i\omega_{n}}
\mathcal{G}_{\downarrow\downarrow}\mathcal{G}_{\downarrow\downarrow}\:\:,\\
\sigma_{\uparrow\downarrow} &=& \beta^{-1}\sum_{i\omega_{n}}
\mathcal{G}_{\uparrow\downarrow}\mathcal{G}_{\downarrow\uparrow}\:\:.
\end{eqnarray}
\noindent By carrying out these Matsubara frequency summations we
get
\begin{eqnarray}
\sigma_{\uparrow\uparrow} &=& \frac{n_{F}(E_{\mathbf{k}}^{+})+
n_{F}(E_{\mathbf{k}}^{-})-1}
{\epsilon_{\Delta}}\:u_{\mathbf{k}}^{2}v_{\mathbf{k}}^{2}+
n_{F}'(E_{\mathbf{k}}^{+})u_{\mathbf{k}}^{4}
+n_{F}'(E_{\mathbf{k}}^{-})v_{\mathbf{k}}^{4}\:\:,\\
\sigma_{\downarrow\downarrow} &=& \frac{n_{F}(E_{\mathbf{k}}^{+})+
n_{F}(E_{\mathbf{k}}^{-})-1}
{\epsilon_{\Delta}}\:u_{\mathbf{k}}^{2}v_{\mathbf{k}}^{2}
+n_{F}'(E_{\mathbf{k}}^{+})
v_{\mathbf{k}}^{4}+n_{F}'(E_{\mathbf{k}}^{-})
u_{\mathbf{k}}^{4} \:\:,\\
\sigma_{\uparrow\downarrow} &=&
\left[\frac{1-n_{F}(E_{\mathbf{k}}^{+})
-n_{F}(E_{\mathbf{k}}^{-})}{\epsilon_{\Delta}}
+n_{F}'(E_{\mathbf{k}}^{+})+n_{F}'(E_{\mathbf{k}}^{-})\right]
u_{\mathbf{k}}^{2}v_{\mathbf{k}}^{2} \:\:,
\end{eqnarray}
\noindent where $n_{F}'(x)\equiv dn_{F}(x)/dx$. Using the last
three equations and identity
$u_{\mathbf{k}}^{2}+v_{\mathbf{k}}^{2}=1$, we then obtain
\begin{equation}\label{rhos}
\rho_{s}=m_{\uparrow}n_{\uparrow}+m_{\downarrow}n_{\downarrow}+
\int\frac{d^{3}{\mathbf k}}{(2\pi)^{3}}\frac{{\mathbf k}^{2}}{3}
\left[n_{F}'(E^{+}_{\mathbf k})+n_{F}'(E^{-}_{\mathbf k})\right]
\:\:.
\end{equation}
\noindent Here $n_{\uparrow}$ and $n_{\downarrow}$ are
momentum-space integrals of $n_{\uparrow}(\mathbf{k})$ and
$n_{\downarrow}(\mathbf{k})$, respectively. In what follows, we
employ formula \eqref{rhos} to determine the superfluid (number)
density $n_{s}=\rho_{s}/(m_{\uparrow}+m_{\downarrow})$ at zero
temperature in the special case of an equal mass system of
interest in the present work.

At zero temperature $n_{F}(x)=\theta(-x)$, whereby for
$m_{\uparrow}=m_{\downarrow}=m$ we readily obtain
\begin{equation}\label{}
\rho_{s}=m(n_{\uparrow}^{0}+n_{\downarrow}^{0})-
\int\frac{d^{3}{\mathbf k}} {(2\pi)^{3}}\frac{{\mathbf
k}^{2}}{3}\left[\delta(E^{-}_{\mathbf k})+\delta(E^{+}_{\mathbf
k})\right]\:\:,
\end{equation}
\noindent where $n_{\uparrow}^{0}$ and $n_{\downarrow}^{0}$ are
the zero-temperature values of $n_{\uparrow}$ and
$n_{\downarrow}$, i.e., the respective momentum-space integrals of
\begin{eqnarray}
n_{\uparrow}^{0}(\mathbf{k}) &=&
u_{\mathbf{k}}^{2}\:\theta(-E^{+}_{\mathbf k})+
v_{\mathbf{k}}^{2}\:\theta(E^{-}_{\mathbf k}) \:\:, \\
n_{\downarrow}^{0}(\mathbf{k}) &=&
u_{\mathbf{k}}^{2}\:\theta(-E^{-}_{\mathbf k})+
v_{\mathbf{k}}^{2}\:\theta(E^{+}_{\mathbf k})\:\:.
\end{eqnarray}
\noindent The squared Bogoliubov amplitudes (coherence factors) in
this special case are given by
\begin{eqnarray}
u_{\mathbf{k}}^{2} &=&
\frac{1}{2}\left(1+\frac{\frac{\mathbf{k}^2}
{2m}-\mu}{\sqrt{(\frac{\mathbf{k}^2}
{2m}-\mu)^2+\Delta^{2}}}\right) \:\:,\\
v_{\mathbf{k}}^{2} &=&
\frac{1}{2}\left(1-\frac{\frac{\mathbf{k}^2}
{2m}-\mu}{\sqrt{(\frac{\mathbf{k}^2}
{2m}-\mu)^2+\Delta^{2}}}\right) \:\:.
\end{eqnarray}
\noindent After performing a trivial angular integration, on
account of the fact that the upper branch $E^{+}_{\mathbf k}$ is
always positive in our case and that $E^{-}_{\mathbf
k}=\varepsilon_{\mathbf k}$, we obtain an expression for the
superfluid density $n_{s}=\rho_{s}/(2m)$ :
\begin{equation}\label{}
n_{s}=\frac{1}{2}(n_{\uparrow}^{0}+n_{\downarrow}^{0})
-\frac{1}{12\pi^{2}m}\int_{0}^{\infty}{|\mathbf
k|}^{4}\delta(\varepsilon_{\mathbf k})\:d{|\mathbf k|}\:\:.
\end{equation}
\noindent We now invoke the property of Dirac's $\delta$ function
\begin{equation}\label{}
 \delta(f(x))=\frac{1}{|f'(x)|}\sum_{i}\delta(x-x_{i}) \:\:,
\end{equation}
\noindent $x_{i}$ being the simple zeros of the function $f(x)$
(i.e., $f(x_{i})=0$, $f'(x_{i})\neq 0$). This property can be
equivalently stated as
\begin{equation}\label{}
\int_{D}h(x)\delta(f(x))dx=\sum_{i}\frac{h(x_{i})}{|f'(x_{i})|}
\:\:,
\end{equation}
\noindent where the last sum extends over all the simple zeros of
$f(x)$ within the domain of integration $D$. On account of the
fact that $|\mathbf k|=k_{b}$ is the only zero of the function
$\varepsilon (|{\mathbf k}|)$, simple transformations lead to
\begin{equation}\label{nseq}
n_{s}=\frac{1}{2}(n_{\uparrow}^{0}+n_{\downarrow}^{0})-
\frac{k_{b}^{3}}{12\pi^{2}}\frac{\sqrt{(\frac{k_{b}^2}
{2m}-\mu)^2+\Delta^{2}}}{\frac{k_{b}^{2}}{2m}-\mu} \:\:.
\end{equation}
\noindent As can straightforwardly be derived, $n_{\uparrow}^{0}$
and $n_{\downarrow}^{0}$ are given by
\begin{eqnarray}
n_{\uparrow}^{0} &=& \frac{1}{2\pi^{2}}\int_{k_{b}}^{\infty}
{|\mathbf k|}^{2} v_{\mathbf{k}}^{2}\:d{|\mathbf k|} \:\:,  \\
n_{\downarrow}^{0} &=&
\frac{1}{2\pi^{2}}\left(\frac{k_{b}^{3}}{3}+ \int_{k_{b}}^{\infty}
{|\mathbf k|}^{2}v_{\mathbf{k}}^{2}\:d{|\mathbf k|}\right) \:\:.
\end{eqnarray}
\noindent Their calculation requires numerical evaluation of the
integral
\begin{equation}\label{}
\int_{k_{b}}^{\infty}{|\mathbf k|}^{2} v_{\mathbf{k}}^{2}
\:d{|\mathbf k|}=\frac{1}{2}\int_{k_{b}}^{\infty} {|\mathbf
k|}^{2}\left(1-\frac{\frac{|\mathbf{k}|^2}
{2m}-\mu}{\sqrt{(\frac{|\mathbf{k}|^2}
{2m}-\mu)^2+\Delta^{2}}}\right)d{|\mathbf k|} \:\:.
\end{equation}
\noindent Finally, we eliminate $k_{b}$ from Eq. \eqref{nseq} (in
favor of parameters $\mu$, $\delta$, and $\Delta$) using the
identity $\sqrt{(\frac{k_{b}^2} {2m}-\mu)^2+\Delta^{2}}=\delta/2$
and thereby obtain :
\begin{equation}\label{}
n_{s}=\frac{1}{2}(n_{\uparrow}^{0}+n_{\downarrow}^{0})-
\frac{\delta}{24\pi^{2}}
\frac{\left\{2m\left(\mu+\sqrt{(\frac{\delta}{2})^{2}
-\Delta^{2}}\right)\right\}^{3/2}}{\sqrt{(\frac{\delta}{2})^{2}
-\Delta^{2}}} \:\:.
\end{equation}

\begin{figure}[htb]
\begin{center}
\includegraphics[width=0.9\linewidth]{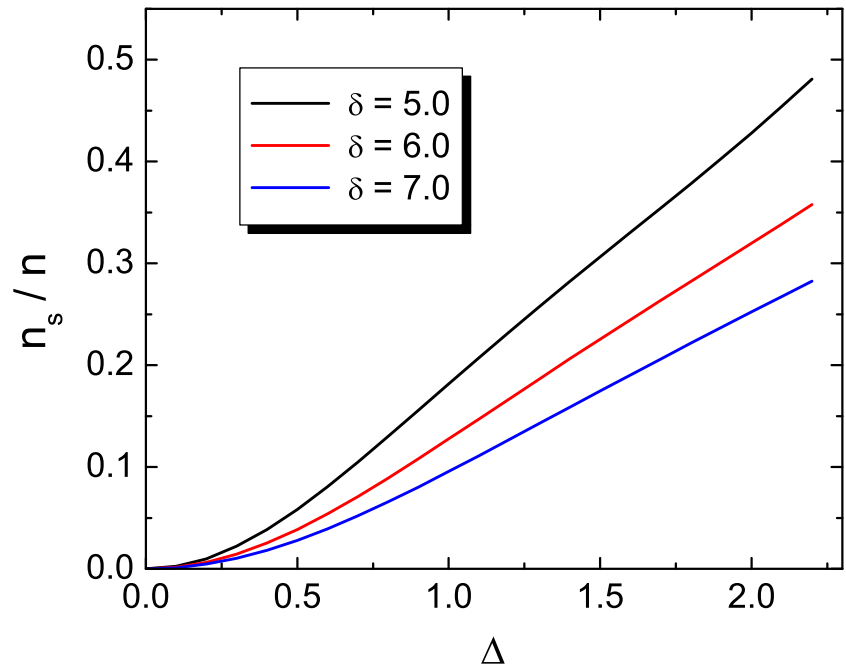}
 \end{center}
\caption{Superfluid density in units of the total atomic density
$n=n_{\downarrow}+n_{\uparrow}$ for three different values of the
chemical potential mismatch $\delta$. Both $\delta$ and $\Delta$
are expressed in units of $|\mu|$.}\label{fig:ns567}
\end{figure}

\begin{figure}
\begin{center}
\includegraphics[width=0.9\linewidth]{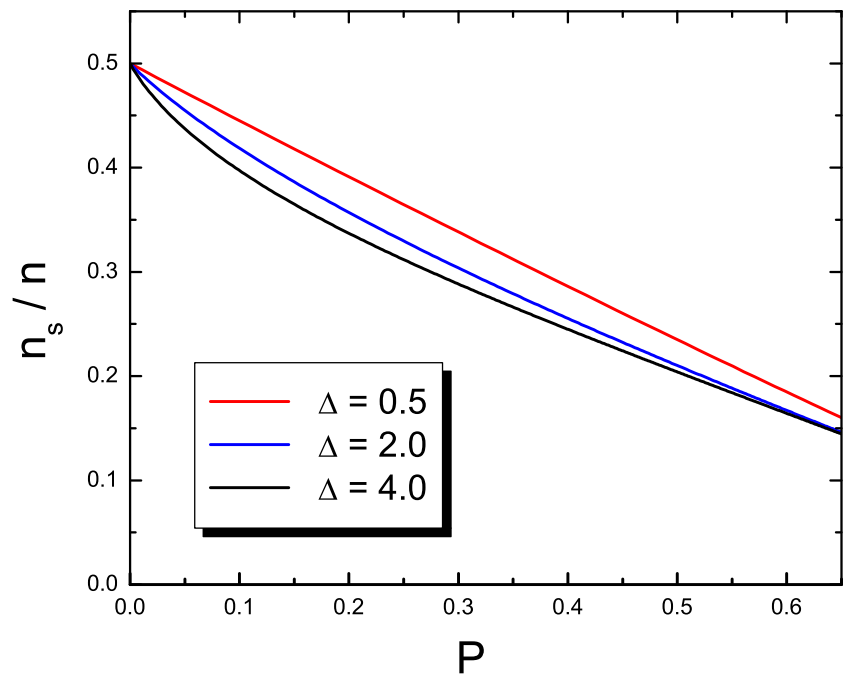}
\end{center}
\caption{Superfluid density as a function of spin polarization for
different values of the pairing gap $\Delta$ (expressed in units
of $|\mu|$).}\label{fig:nsvsP}
\end{figure}

The superfluid density is calculated numerically based on the
derived expressions. Some typical results thereby obtained for the
superfluid density as a function of the pairing gap are presented
in Fig.~\ref{fig:ns567}. In Fig.~\ref{fig:nsvsP} superfluid
density is plotted as a function of the spin-polarization
$P=(n_{\downarrow}-n_{\uparrow})/(n_{\downarrow}+n_{\uparrow})$
for different values of the pairing gap. In contrast to the
related BP2 state, no anomalous negative value of the superfluid
density in the BP1 phase is found, thus corroborating the
dynamical stability of this phase.

\section{Low-energy effective field theory}

In this Section, we start from a symmetry-based low-energy
effective Lagrangian for the gapless branch of fermions and the
superfluid phase field. As is widely accepted, deep in the
superfluid regime the dominant role is played by the superfluid
phase fluctuations, while the fluctuations of the amplitude of the
order parameter can be neglected (the London
limit).~\cite{nelsonbook}  We then derive an effective phase-only
action by integrating out the fermionic degrees of freedom.
The upper cutoff for the wave vector {\bf q} of the phase fluctuations
is set by
$k_{\Delta}=(2m\Delta)^{1/2}$, a momentum scale corresponding to
the pairing gap $\Delta$. This is a consequence of the BP$1$
superfluid phase being realized in the strong-coupling regime on
the BEC side of a Feshbach resonance, where the pairing gap is
related to the binding energy of a Feshbach molecule. In more
common examples of fermionic pairing (e.g., the weak-coupling
regime on the atomic side of a Feshbach resonance) the momentum
cutoff would have been set by $\xi_{0}^{-1}\sim \Delta/v_{F}$ (the
inverse of the coherence length), where
$v_{F}$ is the average Fermi velocity of the pairing fermions. It
is important to point out that the magnitude $|\mathbf{q}|$ of the
wave vector of phase fluctuations is not necessarily small as
compared to $k_{b}$. Namely, as already stated in Sec.~\ref{fn}
(recall Eq.~\eqref{kb}), $k_{b}$ is controlled by the spin
population imbalance and can therefore (by tuning the population
imbalance) be made arbitrarily small.

The form of our effective theory will be chosen so as to obey the
Galilean invariance. In a Galilean invariant system the momentum
density $T_{0i}$ (the off-diagonal part of the stress tensor) has
to be equal to the mass carried by the particle number current
$J_{i}$, i.e.,
\begin{equation}\label{galileicond}
  T_{0i}=mJ_{i}\:\:.
\end{equation}
\noindent This is an example of algebraic identity between
operators implementing symmetries that hold in the microscopic
theory and must be retained in the effective theory.~\cite{gww:89}

\subsection{Parametrization of the effective theory}
\label{sec:parameters}

Throughout the analysis in this work, we shall use $k_{b}$ (or,
alternately, the spin-polarization $P$) and the pairing gap
$\Delta$ as free input parameters that will be compared to
experiments. As to this choice of free parameters, a remark is in
order. In a truly microscopic theory, formulated in terms of the
original fermions, the pairing gap would be determined by solving
the gap equation~\cite{Levinsen:06} together with equations
specifying conservation of the total number of atoms and the
population imbalance. The strength of coupling between fermions,
naturally, shows up in these equations. Our theory, however, is
not microscopic: we here assume the existence of the BP1 phase and
construct an effective field theory for this phase. Being
formulated in terms of collective rather than microscopic degrees
of freedom, our effective theory does not explicitly have the
coupling strength between original fermions and instead uses the
pairing gap as an independent parameter. This choice is also
motivated by the recent experimental developments in the field of
atomic Fermi gases: it was demonstrated that using the
rf-spectroscopy it is possible to measure the pairing gap by
breaking fermion pairs.~\cite{Chin:04,greiner+:05} Alternative
methods of detecting a long-range pairing order in a degenerate
Fermi gas have also been theoretically proposed, where the pairing
function is directly measured in real space via a matter-wave
interferometric techniques.~\cite{Carusotto:05}

\subsection{Symmetry-based effective Lagrangian}

The effective Lagrangian of the system ought to obey two global
$U(1)$ symmetries, one of which corresponds to the total atom
number conservation (to be denoted as $U_{c}(1)$), and the other
one to the conservation of the difference in the number of atoms
of spin-up and spin-down species (denoted as $U_{s}(1)$). Our
low-energy effective Lagrangian for the gapless branch of fermions
(Bogoliubov quasiparticles), described by the field $\chi(x)$, and
the superfluid phase field $\theta(x)$ (in the imaginary-time
path-integral formalism, with $\tau=it$) reads
\begin{equation}\label{lagr}
\mathcal{L}=\chi^{*}[\partial_{\tau}+\varepsilon(-i\nabla)]\chi
+c_{1}(\partial_{\tau}\theta)^{2}+c_{2}(\nabla\theta)^{2}
+c_{3}\chi^{*}\chi\Big[i\partial_{\tau}\theta+\frac{1}{2m_{p}}
(\nabla\theta)^{2}\Big]+\nabla\theta\cdot\mathbf{j}+\ldots\:\:,
\end{equation}
\noindent and represents an extension of the theory derived by Son
and Stephanov~\cite{Son-Stephanov:06} to the case of an arbitrary
spin population imbalance. In \eqref{lagr} the ellipses stand for
possible higher-order derivative terms of the $\theta$ field;
$\mathbf{j}=(\chi^{*}\nabla\chi-\nabla\chi^{*}\chi)/(2m_{p}i)$ is
the ``paramagnetic'' fermion (mass) current with $m_{p}=2m$ being
the total mass of the Cooper pair; $\varepsilon(-i\nabla)$ is the
operator form of the gapless fermion dispersion~\eqref{disp},
written in the coordinate representation. The Lagrangian has the
shift symmetry $\theta\rightarrow \theta+\alpha$, due to the
$U_{c}(1)$ particle number symmetry. Consequently, it contains the
coordinate and time derivatives of $\theta$, but not $\theta$
itself.

The phenomenological parameters $c_{1}$, $c_{2}$ and $c_{3}$ are
not constrained by the $U(1)$ symmetries. While $c_{1}=\partial
n/\partial \mu$ ($n$ being the total atomic
density),~\cite{sonwingate:06,gww:89} $c_{2}$ and $c_{3}$ are
constrained by the superfluid density $n_{s}$.  In this
regard, an important difference between the bosonic (phase) sector
of our theory and the effective low-energy theories of bosonic
superfluids or neutral fully-gapped superconductors (with equal
spin population) ought to be pointed out. Namely, in theories of
the present type, in order to satisfy the Galilean invariance
represented by the constraint \eqref{galileicond}, the low-energy
effective Lagrangian can depend on the phase field only through
the Galilean-invariant combination $U_{\theta}\equiv
\partial_{\tau}\theta+\frac{1}{2m_{0}}(\nabla\theta)^{2}$
($m_{0}$ being the mass of an elementary superfluid constituent,
e.g., the mass of a single atom in the case of $^{4}$He or the
mass of a Cooper pair in case of fermionic superfluids), that is,
\begin{equation}\label{}
\mathcal{L}_{\theta}=P\left(i\partial_{\tau}\theta
+\frac{1}{2m_{0}}(\nabla\theta)^{2}\right) \:\:,
\end{equation}
\noindent where $P$ stands for an arbitrary
polynomial.~\cite{gww:89} Keeping only the terms of the lowest
order in the derivatives of $\theta$, $\mathcal{L}_{\theta}$
reduces to the form $c_{1}(\partial_{\tau}\theta)^{2}
+c_{2}(\nabla\theta)^{2}$, where coefficients $c_{1}$ and $c_{2}$
are fixed by the requirement that this Lagrangian correctly
describes the dynamics of the gapless Goldstone mode
(Anderson-Bogoliubov mode in the case of a neutral superconductor)
associated with the spontaneously broken global $U(1)$ symmetry.
[Note that the term linear in $\partial_{\tau}\theta$ is omitted,
despite being of the lowest order, since it is a total derivative
and the time-dependent topological configurations are not
considered.] In our case, however, with additional low-energy
degrees of freedom (gapless fermion excitations), the coefficient
$c_{2}$ is renormalized at every order of the effective theory and
is constrained together with $c_{3}$ by an additional requirement
that the superfluid density matches the one calculated from the
microscopic theory. This identification will be made in the
following section.

The Galilean invariance of the fermion-dependent part of this
Lagrangian is explicitly demonstrated in Appendix \ref{Galilean}.
An important consequence of this invariance is that the
coefficient of the term $\nabla\theta\cdot \Bj$ must be unity. As
a prerequisite for proving Galilean invariance, we have shown that
the Bogoliubov-quasiparticle field remains invariant under
Galilean transformations. The transformation law for this
quasiparticle field is thus essentially different from that of the
original fermions, used as a basis for an alternative effective
field theory of a polarized Fermi gas in
Ref.~\cite{nishida+son:06}. This is consistent with a quite
general argument that the transformation properties for
quasiparticles in the low-energy effective theories should not
depend on the quantities such as the bare particle mass
$m$.~\cite{VolovikReports:01} As a by-product of this
transformation law, the Bogoliubov-quasiparticle current $\Bj$ is
invariant under Galilean boosts, which is also consistent with the
invariance of the quasiparticle momentum.

\subsection{Effective action for phase fluctuations}\label{effphase}

Using the $\phi$ and {\bf a} fields via Eq.(6), the Lagrangian
\eqref{lagr} can be rewritten as
\begin{equation}\label{lagrang}
\begin{array}{rl}
\mathcal{L}=&\displaystyle
\chi^{*}[\partial_{\tau}+\varepsilon(-i\nabla)]
\chi+c_{1}(\partial_{\tau}\phi)^{2}
+c_{2}(\nabla\phi-\mathbf{a})^{2} \\
&\displaystyle +c_{3}\chi^{*}\chi\Big[i\partial_{\tau}\phi+
\frac{(\nabla\phi-\mathbf{a})^{2}}{2m_{p}}\Big]
+(\nabla\phi-\mathbf{a})\cdot\mathbf{j} \:\:.
\end{array}
\end{equation}

In order to arrive at an effective phase-only action
$S[\theta]\equiv S[\phi,\mathbf{a}]$, we integrate out the fermion
field $\chi$ :
\begin{equation}
 e^{-S[\theta]}=\int D(\chi^{*},\chi)\:e^{-S[\chi,\theta]}\:\:,
\end{equation}
\noindent where $S[\chi,\theta]\equiv
S[\chi,\phi,\mathbf{a}]=\int_{0}^{\beta}d\tau\int
d\mathbf{x}\:\mathcal{L}$ is the Euclidean action corresponding to
Lagrangian \eqref{lagrang}(with $\beta\equiv (k_{B}T)^{-1}$ the
inverse temperature). To this end, we first note that the fermion
field enters Lagrangian \eqref{lagrang} through a quadratic form
$\chi^{*}K\chi=\chi^{*}(-\mathcal{G}_{0}^{-1}+X)\chi$, where
\begin{equation}
\mathcal{G}_{0}=[-\partial_{\tau}-\varepsilon(-i\nabla)]^{-1} \:\:
\end{equation}
\noindent is the noninteracting fermion propagator, and
$X=X^{(1)}+X^{(2)}$ where
\begin{eqnarray}\label{x}
X^{(1)}&=&
i\partial_{\tau}\phi+\frac{1}{2m_{p}i}\:(\nabla\phi-\mathbf{a})\cdot
\dnabla\:\:,\\
X^{(2)}&=& \frac{1}{2m_{p}}\:(\nabla\phi-\mathbf{a})^{2}\:\:,
\end{eqnarray}
\noindent are respectively of the first and second orders in
fields $\phi$ and $\mathbf{a}$.

Integrating out the fermionic degrees of freedom gives rise to a
contribution $S_{F}[\phi,\mathbf{a}]=-\tr\ln K$ to the effective
action $S[\phi,\mathbf{a}]$, where
\begin{equation}\label{eqna}
-\tr\ln K = -\tr \ln(-\mathcal{G}_{0}^{-1})-\tr
\ln(1-\mathcal{G}_{0}X) \:\:.
\end{equation}
\noindent The contribution of the self-energy $X$ to the effective
phase-only action is evaluated by employing the usual
loop-expansion of the trace: by Taylor-expanding the second term
on the right-hand side of the last equation (using
$\ln(1-z)=-\sum_{n=1}^{\infty}z^{n}/n$) we obtain
\begin{equation}
-\tr\ln K = \textrm{const.}+\sum_{n=1}^{\infty}\:\frac{1}{n}
\:\tr[(\mathcal{G}_{0}X)^{n}] \:\:.
\end{equation}
\noindent Using diagonality of the noninteracting fermion
propagator in the momentum-frequency space
($\mathcal{G}_{0}(k,k')\equiv\mathcal{G}_{0}(k)\delta_{kk'}$ with
$\mathcal{G}_{0}(k)=(i\omega_{n}-\varepsilon_{\mathbf{k}})^{-1}$,
here displayed using compact four-momentum notation : $k
\equiv(\mathbf{k},i\omega_{n})$), it is straightforward to show
that
\begin{eqnarray}
\tr(\mathcal{G}_{0}X)&=&\frac{1}{\beta
V}\sum_{k}\mathcal{G}_{0}(k)X_{k,k}\:\:, \label{fir} \\
\tr[(\mathcal{G}_{0}X)^{2}]&=&\frac{1}{(\beta
V)^{2}}\sum_{k,q}\mathcal{G}_{0}(k)X_{k,k+q}\mathcal{G}_{0}(k+q)
X_{k+q,k}\:\:,\label{sec}
\end{eqnarray}
\noindent where $X_{k,k'}$ stands for the Fourier transform of
$X$. In order to obtain the effective action $S[\phi,\mathbf{a}]$
to second order in fields $\phi$ and $\mathbf{a}$, we employ the
above expansion to first order in $X^{(2)}$ (tree level) and to
second order in $X^{(1)}$ (one-loop order).

The tree-level contribution of $X^{(2)}$ to
$S_{F}[\phi,\mathbf{a}]$ (and therefore to the effective
phase-only action) is obtained by replacing $\chi^{*}\chi$ by its
average value $\langle\chi^{*}\chi\rangle=n_{b}$. It can easily be
demonstrated that
\begin{equation}
X^{(1)}_{k,k'}=(\omega_{n}-\omega_{n'})\phi_{k-k'}
+\frac{1}{2m_{p}}(\mathbf{k}+\mathbf{k'})\cdot
\left\{\mathbf{a}_{k-k'}-i(\mathbf{k}-\mathbf{k'})\phi_{k-k'}
\right\}\:\:,
\end{equation}
\noindent where $\mathbf{a}_{k}\equiv
\mathbf{a}_{\mathbf{k}}\delta_{\omega_{n},0}$ (the vortex gauge
field is time-independent, i.e. classical). As a special case of
the last equation, in the previously adopted Coulomb gauge (in
which $\mathbf{q}\cdot \mathbf{a}_{\pm \mathbf{q}} =0$, hence
$\mathbf{q}\cdot \mathbf{a}_{\pm q} =0$) we obtain
\begin{eqnarray}
X^{(1)}_{k+q,k} &=& - \left\{\omega_{l}-\frac{i}{m_{p}}\:
\mathbf{q}\cdot\left(\mathbf{k}+\frac{\mathbf{q}}{2}\right)\right\}
\phi_{q}-\frac{1}{m_{p}}\:\mathbf{k}_{\perp}\cdot \mathbf{a}_{q}
\:\:,\\
X^{(1)}_{k,k+q} &=& \left\{\omega_{l}-\frac{i}{m_{p}}\:\mathbf{q}
\cdot\left(\mathbf{k}+\frac{\mathbf{q}}{2}\right)\right\}
\phi_{-q}-\frac{1}{m_{p}}\:\mathbf{k}_{\perp}\cdot \mathbf{a}_{-q}
\:\:,
\end{eqnarray}
\noindent where $q \equiv (\mathbf{q},i\omega_{l})$ and
$\mathbf{k}_{\perp}\equiv\{(\mathbf{q}\times \mathbf{k}) \times
\mathbf{q}\}/\mathbf{q}^{2}$ is the transverse component of the
three-dimensional vector $\mathbf{k}$ with respect to
$\mathbf{q}$. While it is easy to show that the first order
contribution (tree level) of $X^{(1)}$ is equal to zero, by
inserting the last two equations into Eq. \eqref{sec} we find its
contribution to $S_{F}[\phi,\mathbf{a}]$ at one-loop order.

The effective action for $\phi$ and $\mathbf{a}$ is obtained by
gathering $S_{F}[\phi,\mathbf{a}]$ and the fermion-independent
terms of the original action :
\begin{equation}
S[\phi,\mathbf{a}]=\int_{0}^{\beta}d\tau\int
d\mathbf{x}\left[c_{1}(\partial_{\tau}\phi)^{2}+c_{2}
(\nabla\phi-\mathbf{a})^{2}\right]+S_{F}[\phi,\mathbf{a}]\:\:.
\end{equation}
\noindent  In the momentum-frequency space, to second order in
fields $\phi$ and $\mathbf{a}$, it is represented by the quadratic
form
\begin{equation}\label{phasefl}
\begin{array}{rl}
S[\phi,\mathbf{a}]=& \displaystyle\sum_{q}\left\{
\left(c_{2}+\frac{n_{b}}{2m_{p}}\:c_{3}\right)
\mathbf{q}^{2}+\frac{1}{2m_{p}^{2}}R_{ij}(q)q_{i}q_{j}
+\left(c_{1}-\frac{\Pi(q)}{2}\right)
\omega_{l}^{2}\right\}\phi_{q}\phi_{-q} \\
&\displaystyle+\sum_{q}\left\{c_{2}+\frac{n_{b}}{2m_{p}}\:c_{3}
+\frac{P(q)}{2m_{p}^{2}}\right\}{\mathbf{a}}_{q}
\cdot{\mathbf{a}}_{-q} \:\:\:.
\end{array}
\end{equation}
\noindent The first term corresponds to the propagating
Goldstone modes of broken $U(1)$ symmetry, and the second one to
its corresponding topological defects - vortices. [Summation over
repeated indices in the last equation is implicit.] Here
\begin{equation}
\Pi(q) = \frac{1}{\beta V}
\sum_{k}\mathcal{G}_{0}(k)\mathcal{G}_{0}(k+q)
\end{equation}
\noindent is the fermion density polarization bubble, while
\begin{eqnarray}
R_{ij}(q) &=& \frac{1}{\beta V}
\sum_{k}\mathcal{G}_{0}(k)\mathcal{G}_{0}(k+q)
\left(k_{i}+\frac{q_{i}}{2}\right)\left(k_{j}
+\frac{q_{j}}{2}\right) \:\:, \\
P(q) &=& \frac{1}{\beta V}\sum_{k}\mathcal{G}_{0}(k)
\mathcal{G}_{0}(k+q)\:\frac{\mathbf{k}_{\perp}^{2}}{2}
\label{pqdef}
\end{eqnarray}
\noindent represent the longitudinal and transverse
current-current correlation functions, respectively. In obtaining
the form of the latter, we have made use of the identity
\begin{equation}\label{}
\sum_\mathbf{k}k_{\perp i} k_{\perp j}
F(|\mathbf{k}|)=\frac{\delta_{ij}}{2}\sum_\mathbf{k}
\mathbf{k}_{\perp}^{2}F(|\mathbf{k}|) \:\:,
\end{equation}
\noindent valid for any rotationally-invariant function
$F(|\mathbf{k}|)$. The Matsubara frequency sum that is implicit in
all of these response functions evaluates to
\begin{equation}\label{}
\frac{1}{\beta}
\sum_{i\omega_{n}}\mathcal{G}_{0}(k)\mathcal{G}_{0}(k+q)=
\frac{n_{F}(\varepsilon_{\mathbf{k}})
-n_{F}(\varepsilon_{\mathbf{k+q}})}
{i\omega_{l}+\varepsilon_{\mathbf{k}}-\varepsilon_{\mathbf{k+q}}}
\:\:.
\end{equation}

It is important to point out that there is no RPA-type correction
from the interaction vertex $\Bj \cdot \nabla\phi$ to the
transverse current-current correlation function; this is manifest
in our choice of the Coulomb gauge for the topological gauge field
${\bf a}$.

As proven in Appendix \ref{rqocalc}, $R_{ij}(q)= R(q)\delta_{ij}$.
Consequently, the phase-only action in Eq. \eqref{phasefl} in the
zero-temperature static limit reduces to
\begin{equation}\label{phasef2}
\begin{array}{rl}
S[\phi,\mathbf{a}]=& \displaystyle\sum_{\mathbf{q}}
\left(c_{2}+\frac{n_{b}}{2m_{p}}\:c_{3}
+\frac{R^{0}_{\mathbf{q}}}{2m_{p}^{2}}\right)\mathbf{q}^{2}
\phi_{\mathbf{q}}\phi_{-\mathbf{q}} \\
&\displaystyle+\sum_{\mathbf{q}}\left(c_{2}+\frac{n_{b}}{2m_{p}}\:c_{3}
+\frac{P^{0}_{\mathbf{q}}}{2m_{p}^{2}}\right)
{\mathbf{a}}_{\mathbf{q}}\cdot{\mathbf{a}}_{-\mathbf{q}} \:\:,
\end{array}
\end{equation}
\noindent where $P^{0}_{\mathbf{q}}$ and $R^{0}_{\mathbf{q}}$ are
the zero-temperature static limits of $P(q)$ and $R(q)$,
respectively.

The superfluid mass density $\rho_{s}$, which plays the role of
rigidity in the present problem (``spin-wave'' stiffness in the
XY-model terminology),~\cite{chaikinlub} can be identified from
the long-wavelength ($\mathbf{q}\rightarrow 0$) limit through the
relation
\begin{equation}\label{}
\frac{\rho_{s}}{2}=c_{2}+\frac{n_{b}}{2m_{p}}\:c_{3}+
\frac{R^{0}_{\mathbf{q}=0}}{2m_{p}^{2}} \:\:.
\end{equation}
\noindent This constraint on $c_{2}$ and $c_{3}$ can be
equivalently stated as
\begin{equation}\label{}
c_{2}+\frac{n_{b}}{2m_{p}}\:c_{3}=\frac{n_{s}}{2m_{p}}-
\frac{R^{0}_{\mathbf{q}=0}}{2m_{p}^{2}}
\end{equation}
\noindent and implies that the phase-only action in Eq.
\eqref{phasef2} adopts the form
\begin{equation}\label{phasef3}
S[\phi,\mathbf{a}]=\sum_{\mathbf{q}}
\frac{n_{s}}{2m_{p}}\:\mathbf{q}^{2}\phi_{\mathbf{q}}
\phi_{-\mathbf{q}}+\sum_{\mathbf{q}}\left(\frac{n_{s}}{2m_{p}}
+\frac{P^{0}_{\mathbf{q}}-R^{0}_{\mathbf{q}=0}}{2m_{p}^{2}}
\right){\mathbf{a}}_{\mathbf{q}}\cdot{\mathbf{a}}_{-\mathbf{q}}
\:\:,
\end{equation}
\noindent which is free of the phenomenological parameters of the
original theory.

\section{Effective theory for vortices and the interaction potential}
\label{vortef}

Starting from the effective theory of phase fluctuations
(described by \eqref{phasef3}) and integrating out the regular
(spin-wave) part of the phase field, we derive the effective
action $S_{\mathrm{eff}}[\mathbf{a}]$ for vortices :
\begin{equation}\label{}
e^{-S_{\mathrm{eff}}[\mathbf{a}]}=\int D(\bar{\phi},\phi)\:
e^{-S[\phi,\mathbf{a}]}\:\:.
\end{equation}
\noindent Along these lines, a straightforward Gaussian functional
integration yields the result
\begin{equation}\label{rezultat}
S_{\mathrm{eff}}[\mathbf{a}]=\sum_{\mathbf{q}}\left(\frac{n_{s}}{2m_{p}}
+\frac{P^{0}_{\mathbf{q}}-R^{0}_{\mathbf{q}=0}}{2m_{p}^{2}}
\right)\mathbf{a}_{\mathbf{q}}\cdot \mathbf{a}_{\mathbf{-q}}\:\:.
\end{equation}
\noindent (Because the vortex gauge field belongs to the classical
sector of the theory, the derived effective action contains only
the $\omega_{l}=0$ part). With the aid of identity \eqref{ind},
the last result can be conveniently recast as
\begin{equation}\label{sef}
S_{\mathrm{eff}}=\sum_{\mathbf{q}}\tilde{\rho}(\mathbf{q})
\kappa_{0}^{2}\left\{\frac{n_{s}}{2m_{p}}
\frac{1}{\mathbf{q}^{2}}+\frac{1}{2m_{p}^{2}}
\frac{P^{0}_{\mathbf{q}}-R^{0}_{\mathbf{q}=0}}
{\mathbf{q}^{2}}\right\}\tilde{\rho}(\mathbf{-q})\:\:.
\end{equation}

From the last equation we read off the momentum-space form of the
effective interaction potential between the vortices :
\begin{equation} \label{effpot}
V_{\mathrm{eff}}(\mathbf{q})=\kappa_{0}^{2}
\left(\frac{n_{s}}{2m_{p}}\frac{1}{\mathbf{q}^{2}}
+\frac{1}{2m_{p}^{2}}\frac{P^{0}_{\mathbf{q}}
-R^{0}_{\mathbf{q}=0}}{\mathbf{q}^{2}}\right) \:\:.
\end{equation}
\noindent  In addition to the long-range component
proportional to $1/\mathbf{q}^{2}$ (logarithmic interaction in the
real space, i.e., $2$D Coulomb potential), characteristic of the
conventional two-dimensional charge-neutral superfluids, we have
an additional component
\begin{equation}\label{vind}
 V_{\mathrm{ind}}(\mathbf{q})=\frac{\kappa_{0}^{2}}{2m_{p}^{2}}
 \frac{P^{0}_{\mathbf{q}}-R^{0}_{\mathbf{q}=0}}{\mathbf{q}^{2}}
\end{equation}
\noindent due to the presence of gapless fermions.

\subsection{Properties of $P_{\mathbf{q}}^{0}$ and $R_{\mathbf{q}}^{0}$}

To calculate $P^{0}_{\mathbf{q}}$ one has to resort to a numerical
evaluation. Yet, before embarking on numerical work we can put
$P^{0}_{\mathbf{q}}$ into a convenient analytical form. In
Appendix~\ref{pqo} we demonstrate that $P^{0}_{\mathbf{q}}$ can be
reduced to a two-dimensional principal-value integral
\begin{equation} \label{simpform}
P^{0}_{\mathbf{q}}=\frac{k_{b}^{3}}{(2\pi)^{2}}\mathcal{P}
\int_{0}^{1}|\mathbf{k}|^{4}d|\mathbf{k}| \int_{-1}^{1}dx
\frac{1-x^{2}}{\sqrt{\xi_{\mathbf{k}}^{2}+
\left(\frac{\Delta}{k_{b}^2}\right)^{2}}
-\sqrt{\left(\xi_{\mathbf{k}}+\frac{|\mathbf{q}|^2}{2m}+
\frac{|\mathbf{k}||\mathbf{q}|}
{m}x\right)^2+\left(\frac{\Delta}{k_{b}^2}\right)^{2}}}\:\:,
\end{equation}
\noindent where momenta $\mathbf{k}$ and $\mathbf{q}$ are
expressed in units of $k_{b}$ and
$\xi_{\mathbf{k}}\equiv|\mathbf{k}|^{2}/2m-(\mu/k_{b}^{2})$. The
presence of the prefactor $k_{b}^{3}\propto
(n_{\downarrow}-n_{\uparrow})$ indicates that in the thermodynamic
limit the induced potential is proportional to the density of
gapless fermions, which could have been expected on physical
grounds.

In the regime of small $|\mathbf{q}|$ ($|\mathbf{q}|<0.1
k_{\Delta}$) numerical evaluation becomes rather troublesome due
to the strongly singular character of the integrand in
Eq.~\eqref{simpform}. However, as demonstrated in Appendix
\ref{anres}, by replacing dispersion $\varepsilon_{\mathbf{k}}$
with its linearized form ($\varepsilon_{\mathbf{k}}\rightarrow
v_{b}(|\mathbf{k}|-k_{b})$), for $|\mathbf{q}|\ll k_{b}$ we can
derive an expression for $P^{0}_{\mathbf{q}}$ in the form of a
controlled expansion in powers of $|\mathbf{q}|/k_{b}$ :
\begin{equation}\label{poqexp}
P^{0}_{\mathbf{q}}= -\frac{k_{b}^{4}}{6\pi^{2}v_{b}}
-\frac{k_{b}^{4}}{10\pi^{2}v_{b}}
\left(\frac{|\mathbf{q}|}{k_{b}}\right)^{2}
+\mathcal{O}\left(\frac{|\mathbf{q}|^{4}}{k_{b}^{4}}\right)\qquad
(\: |\mathbf{q}| \ll k_{b}\:) \:\:.
\end{equation}
\noindent Thus in the $|\mathbf{q}|\rightarrow 0$ limit we obtain:
\begin{equation}\label{poqform}
 P^{0}_{\mathbf{q}} \rightarrow -\frac{k_{b}^{4}}{6\pi^{2}v_{b}}
 \qquad (\: |\mathbf{q}|\rightarrow 0 \:) \:\:.
\end{equation}
\noindent The last result can be given in a more concrete form.
Applying the general expression
\begin{equation}\label{}
 v_{b}=\left|\frac{\partial\varepsilon_{\mathbf k}}
 {\partial\mathbf{k}}\right|_{|{\mathbf k}|=k_{b}}\:\:,
\end{equation}
\noindent to  the case of dispersion \eqref{disp}, we find
\begin{equation}\label{}
 v_{b}=\frac{k_{b}}{m}\frac{\frac{k_{b}^{2}}{2m}-\mu}
 {\sqrt{(\frac{k_{b}^2}{2m}-\mu)^2+\Delta^{2}}}\:\:.
\end{equation}
\noindent Inserting the last result into Eq.~\eqref{poqform} gives
\begin{equation}\label{pqozero}
P^{0}_{\mathbf{q}} \rightarrow -\frac{mk_{b}^{3}}{6\pi^{2}}
\frac{\sqrt{(\frac{k_{b}^2}{2m}-\mu)^2+\Delta^{2}}}
{\frac{k_{b}^{2}}{2m}-\mu} \qquad (\: |\mathbf{q}|\rightarrow 0
\:) \:\:.
\end{equation}

\begin{figure}
\begin{center}
\includegraphics[width=0.9\linewidth]{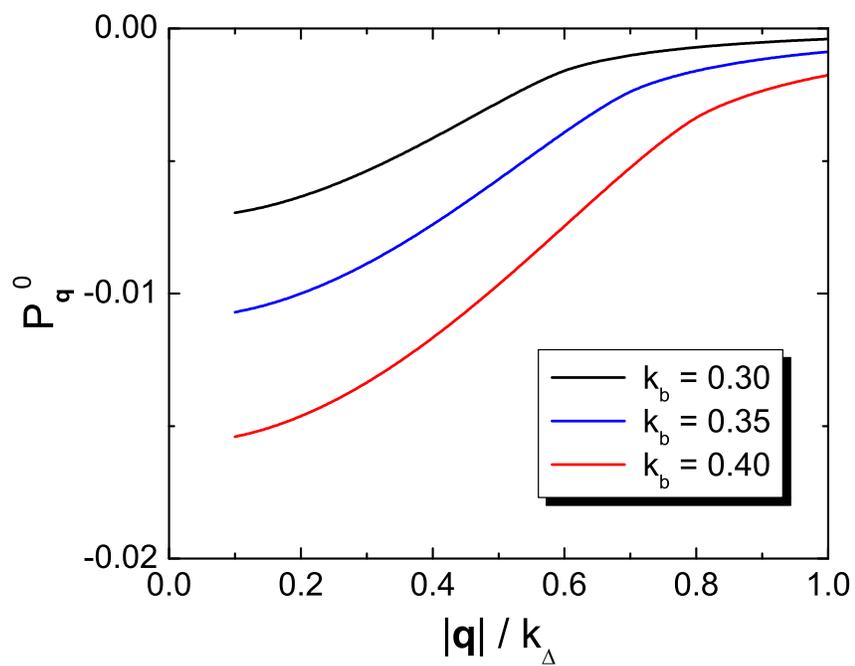}
\end{center}
\caption{Transverse current response function $P_{\mathbf{q}}^{0}$
as a function of dimensionless momentum, for $m=1.0$ and
$\Delta/|\mu|=2.0$. Values of $k_{b}$ are given in units of
$k_{\Delta}$.}\label{fig:pq0}
\end{figure}

Some typical results of numerical evaluation of the response
function $P^{0}_{\mathbf{q}}$ for $0.1 k_{\Delta}\leq
|\mathbf{q}|\leq k_{\Delta}$ are displayed in Fig. \ref{fig:pq0}
(where $2k_{b}<k_{\Delta}$). The salient characteristic of these
results is a knee-like feature at $|\mathbf{q}|=2k_{b}$, which
reflects the existence of an effective Fermi surface with diameter
$2k_{b}$. It bears analogy to the $2k_{F}$-feature of the
paramagnetic spin susceptibility in $3$D, responsible for the RKKY
indirect-exchange interaction between magnetic impurities in
non-magnetic metals,~\cite{RKKY} albeit the $2k_b$-feature found
here comes from the current-current correlator so that it is
different from the RKKY interaction in its dynamical origin. The
values of $P^{0}_{\mathbf{q}}$ obtained analytically in
$|\mathbf{q}|\rightarrow 0$ limit, based on Eq. \eqref{pqozero},
differ just slightly from numerical values at $|\mathbf{q}|= 0.1
k_{\Delta}$, indicating that $P^{0}_{\mathbf{q}}$ can be
approximated as a constant in this numerically-inaccessible region
$0<|\mathbf{q}|< 0.1 k_{\Delta}$. The fact that
$P^{0}_{\mathbf{q}}$ has very weak momentum dependence at small
$\mathbf{q}$ can be inferred from the coefficients in the
controlled expansion of $P^{0}_{\mathbf{q}}$ given by Eq.
\eqref{poqexp}.

In Appendix \ref{rqocalc}, using methodology analogous to the
one employed in Appendix \ref{anres}, we show that
$|\mathbf{q}|\rightarrow 0$ limit of $R^{0}_{\mathbf{q}}$ is equal
to that of $P^{0}_{\mathbf{q}}$ :
\begin{equation}\label{}
R^{0}_{\mathbf{q}=0}=P^{0}_{\mathbf{q}=0}=
-\frac{k_{b}^{4}}{6\pi^{2}v_{b}} \:\:,
\end{equation}
\noindent whereby Eq. \eqref{vind} can be recast as
\begin{equation}\label{vind2}
V_{\mathrm{ind}}(\mathbf{q})=\frac{\kappa_{0}^{2}}{2m_{p}^{2}}
\frac{P^{0}_{\mathbf{q}}-P^{0}_{\mathbf{q}=0}}{\mathbf{q}^{2}}
\:\:.
\end{equation}
\noindent Now, by virtue of controlled expansion \eqref{poqexp},
we obtain that
\begin{equation}\label{}
V_{\mathrm{ind}}(\mathbf{q})=
-\frac{\kappa_{0}^{2}}{2m_{p}^{2}}\frac{k_{b}^{2}}{10\pi^{2}v_{b}}
+\mathcal{O}\left(\frac{|\mathbf{q}|^{2}}{k_{b}^{2}}\right)\qquad
(\: |\mathbf{q}| \ll k_{b}\:) \:\:,
\end{equation}
\noindent and, in particular,
\begin{equation}\label{}
V_\mathrm{ind}(\mathbf{q}=0)=\int
V_{\textrm{ind}}(r)\:d^{2}\mathbf{r}=2\pi\int_{0}^{\infty}
rV_{\textrm{ind}}(r)\:dr
\end{equation}
\noindent is finite :
\begin{equation}\label{}
V_\mathrm{ind}(\mathbf{q}=0)= -\frac{\kappa_{0}^{2}}{2m_{p}^{2}}
\frac{k_{b}^{2}}{10\pi^{2}v_{b}} \:\:.
\end{equation}

\subsection{Effective vortex interaction potential in real space}

Let $F(|\mathbf{q}|)$ be a rotationally-invariant function in
momentum space and $\Lambda$ the upper momentum cutoff. The
inverse two-dimensional Fourier transform of $F(|\mathbf{q}|)$ is
given by
\begin{equation}\label{invft}
 F(r)=\frac{1}{(2\pi)^{2}}\int_{|\mathbf{q}|\leq \Lambda}
 F(|\mathbf{q}|)\:e^{i\mathbf{q}\cdot\mathbf{r}}\:d^{2}\mathbf{q} \:\:.
\end{equation}
\noindent Using the identity
\begin{equation}\label{}
 \int_{0}^{2\pi}e^{i|\mathbf{q}|r\cos\varphi}d\varphi =
 2\pi J_{0}(|\mathbf{q}|r) \:\:,
\end{equation}
\noindent where $J_{0}(x)$ is the zeroth-order Bessel function of
the first kind, the last equation becomes
\begin{equation}\label{}
F(r)=\frac{1}{2\pi}\int_{0}^{\Lambda}|\mathbf{q}|
F(|\mathbf{q}|)J_{0}(|\mathbf{q}|r)\:d|\mathbf{q}| \:\:.
\end{equation}

In our effective theory, the upper momentum cutoff is set by
$k_{\Delta}$, thus the induced potential in real space is given by
\begin{equation}\label{}
V_{\mathrm{ind}}(r)=\frac{1}{2\pi}\int_{0}^{k_{\Delta}}
|\mathbf{q}|V_{\mathrm{ind}}(|\mathbf{q}|)
J_{0}(|\mathbf{q}|r)\:d|\mathbf{q}| \:\:,
\end{equation}
\noindent viz.,
\begin{equation}\label{}
V_{\mathrm{ind}}(r)=\frac{\kappa_{0}^{2}}{4\pi m_{p}^{2}}
\int_{0}^{k_{\Delta}}\frac{P^{0}_{\mathbf{q}}
-P^{0}_{\mathbf{q}=0}}{|\mathbf{q}|}
J_{0}(|\mathbf{q}|r)\:d|\mathbf{q}| \:\:.
\end{equation}

\begin{figure}
\begin{center}
\includegraphics[width=0.9\linewidth]{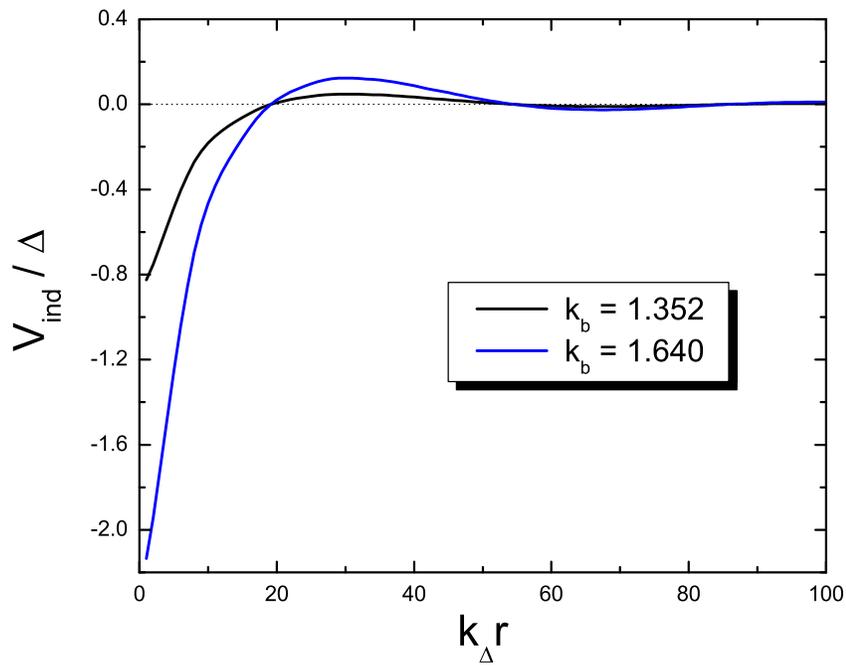}
\end{center}
\caption{Induced vortex interaction potential in real space (in
units of $\Delta$), for $m=1.0$ and $\Delta/|\mu|=1.0$. Values of
$k_{b}$ are given in units of $k_{\Delta}$: $k_{b}=1.352;1.640$
correspond to polarizations $P=0.702; 0.808$,
respectively.}\label{fig:inducedonly}
\end{figure}

Our numerical calculations of $V_{\mathrm{ind}}(r)$ for
different values of relevant parameters ($k_{b},\Delta$) show that
the induced potential has damped oscillatory character, closely
resembling the spatial dependence of the RKKY exchange integral.
As can be seen from Fig.~\ref{fig:inducedonly} this induced
potential has alternating attractive ($dV_\mathrm{ind}/dr>0$) and
repulsive ($dV_\mathrm{ind}/dr<0$) parts. At short distances the
induced potential is always attractive, and the first repulsive
branch appears at the length scale $r\sim (10-25)k_{\Delta}^{-1}$,
depending on the polarization. Spatial period of the observed
oscillations is set by the spin polarization, but is not so simply
related to the radius of the effective Fermi surface as in the
case of genuine RKKY or Friedel oscillations.

The total (effective) vortex-vortex interaction potential in real
space is given by the sum of the induced potential and the
conventional repulsive logarithmic potential. The latter is given
by
\begin{equation}\label{}
 V_{0}(r)=-\kappa_{0}^{2}\frac{n_{s}}{2m_{p}}\ln(k_{\Delta}r)
 \:\:,
\end{equation}
\noindent where the superfluid density $n_{s}$ is calculated in
Sec. \ref{sdcalc}. As our calculations demonstrate, the
effective vortex-vortex interaction shows three characteristic
types of behavior, i.e. three polarization-dependent regimes. The
critical polarizations corresponding to the boundaries between
these different regimes are not universal but depend on the actual
location in the part of the phase diagram pertaining to the BP1
phase.

\begin{figure}
\begin{center}
\includegraphics[width=0.9\linewidth]{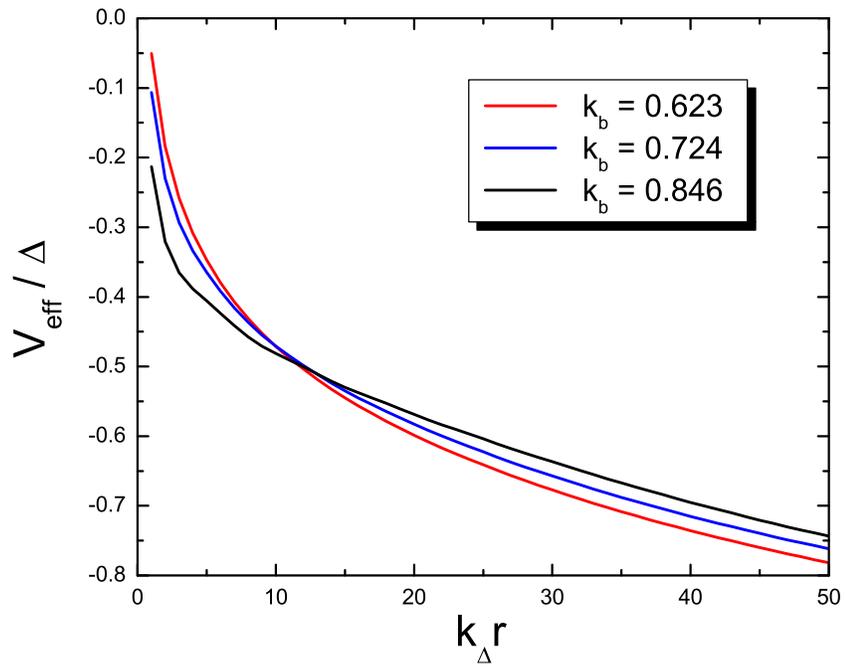}
\end{center}
\caption{Effective vortex interaction potential in real space (in
units of $\Delta$), for $m=1.0$ and $\Delta/|\mu|=2.0$. Values of
$k_{b}$ are given in units of $k_{\Delta}$: $k_{b}$ = 0.623;
0.724; 0.826 correspond to polarizations $P$ = 0.155; 0.227;
0.314, respectively.}\label{fig:effpot1}
\end{figure}

In the regime of relatively low polarization, the total potential
is dominated by the conventional repulsive logarithmic part; the
effective vortex interaction is repulsive
($dV_{\textrm{eff}}/dr<0$) at all distances. The resulting vortex
phase is accordingly expected to be conventional, with triangular
vortex arrangement. An example is shown in
Fig.~\ref{fig:effpot1}.

\begin{figure}
\begin{center}
\includegraphics[width=0.9\linewidth]{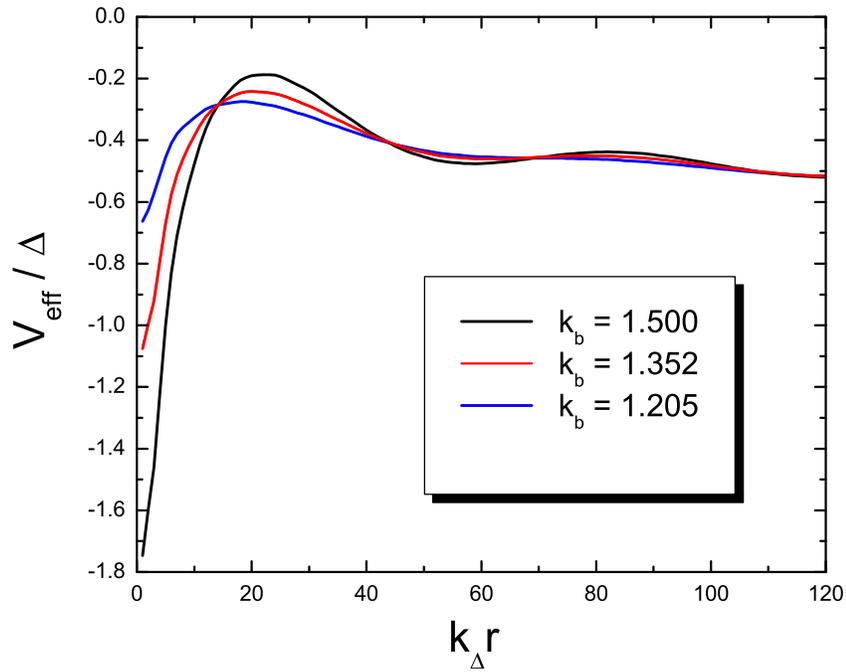}
\end{center}
\caption{Effective vortex interaction potential in real space (in
units of $\Delta$), for $m=1.0$ and $\Delta/|\mu|=1.0$. Values of
$k_{b}$ are given in units of $k_{\Delta}$: $k_{b}$ = 1.500;
1.352; 1.205 correspond to polarizations $P$ = 0.763; 0.702;
0.626, respectively.}\label{fig:effpot2}
\end{figure}

\begin{figure}
\begin{center}
\includegraphics[width=0.9\linewidth]{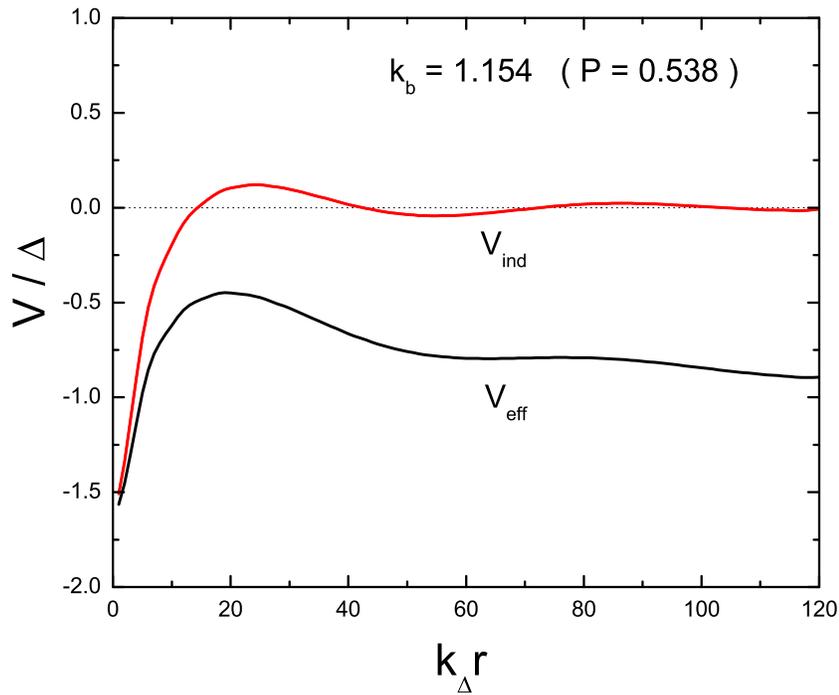}
\end{center}
\caption{Comparison of the induced and the effective vortex
interaction potential in real space (in units of $\Delta$), for
$m=1.0$, $\Delta/|\mu|=2.0$. Values of $k_{b}$ (in units of
$k_{\Delta}$) and $P$ are indicated in the
plot.}\label{fig:indeffpot1}
\end{figure}

In the other extreme - the regime of high polarization, the
induced potential plays a dominant role at short and intermediate
distances. This renders the total potential attractive at short
distances, with pronounced oscillating features resembling the
RKKY interaction, as illustrated in Fig.~\ref{fig:effpot2}.
Comparison of the induced and the total vortex-vortex interaction
potential in the high polarization regime is depicted in
Fig.~\ref{fig:indeffpot1}.

The attractive nature of two-body interaction already at short
distances suggests an instability of the vortex lattice. However,
whether this instability really occurs is still an open question
for the following reasons. The physics at distances shorter than
the healing length $\xi$ (to be discussed in the next section) is
not captured by our effective theory ; also, the multi-vortex
interactions, not considered here but certainly allowed as higher
orders in the effective vortex action, may support unusual vortex
phases. This regime thus requires more elaborate further
investigation.

\begin{figure}
\begin{center}
\includegraphics[width=0.9\linewidth]{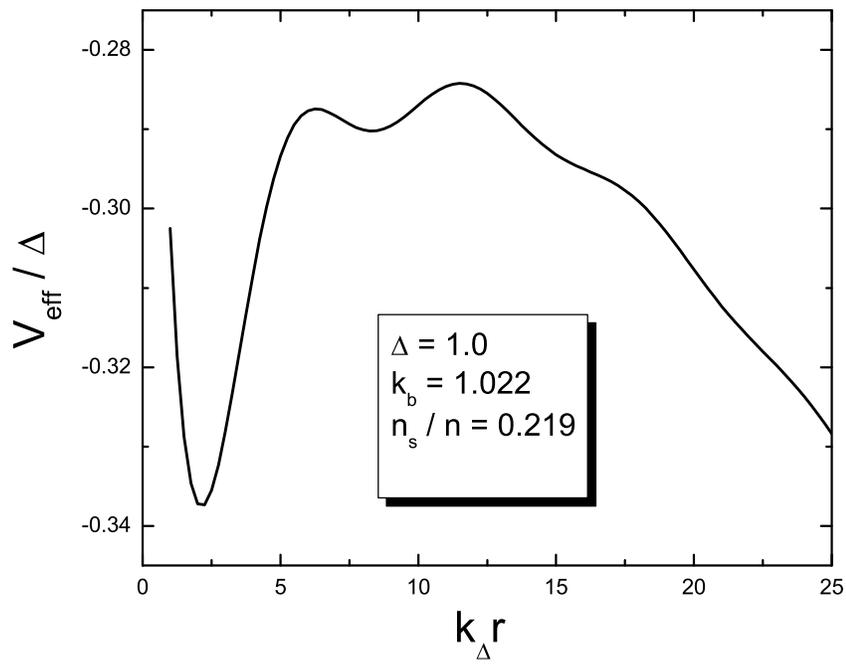}
\end{center}
\caption{Effective vortex interaction potential in real space (in
units of $\Delta$), for $m=1.0$ and $\Delta/|\mu|=1.0$.
$k_{b}=1.022$ (in units of $k_{\Delta}$) corresponds to
$P=0.504$.}\label{fig:effpot3}
\end{figure}

Apart from the two extreme regimes already described, in a narrow
window of parameters the total potential is repulsive at short
distances ($r\approx (2-3)k_{\Delta}^{-1}$) and becomes attractive
at intermediate ones. This intermediate regime is illustrated in
Fig.~\ref{fig:effpot3}.

Due to the finite range of the RKKY-like induced potential, the
truly long-distance dependence of the effective potential is
governed by the infinite-range repulsive logarithmic interaction.
However, for sufficiently large polarization, the effective
potential is non-monotonous function of the distance between two
vortices, a behavior that could potentially give rise to some
exotic vortex-lattice structure. As is well known, the
triangular-lattice configuration minimizes the energy of a system
of point vortices interacting through a repulsive logarithmic
($2$D Coulomb) potential.~\cite{campbelldoria:89}Physically, this
is a consequence of the fact that triangular vortex arrangement
provides maximum nearest-neighbor distance at fixed vortex density
per unit area, which is a natural tendency with purely repulsive
interactions (at least in the continuum, i.e., in the absence of a
vortex-pinning lattice structure). Examples of such behavior can
be found even in physical situations unrelated to vortices, such
as the low-density limit of an electron gas, where a triangular
Wigner crystal is formed. Interestingly, as can be inferred from
Ref.~\cite{campbelldoria:89}, with the conventional logarithmic
interaction the total energy of the triangular configuration of
point-vortices is only around $0.8\%$ smaller than that of the
square-lattice configuration. Such a small difference, however, is
not very surprising given that the lattice periods of these two
configurations (for the same aerial vortex densities) are also not
very different, namely
$a_{\textrm{tr}}=\sqrt{2/\sqrt{3}}\:a_{\textrm{sq}}\simeq
1.0746\:a_{\textrm{sq}}$. For our modified potential between
vortices, which is not repulsive at all distance scales, the
structure of the vortex lattice is an open issue. It is worth
mentioning that a non-monotonous interaction potential between
vortices (albeit without oscillating character) has recently been
found in multicomponent superconductors by Babaev and
Speight.~\cite{Babaev:05} The authors have also predicted the
existence of exotic (non-triangular) vortex-lattice structures.

In general, the interactions between topological defects
mediated by the environment in which they are embedded  is an
important subject of current interest.
Very interesting in
this regard was the study of nodal-quasiparticle-induced
interaction between vortices in d-wave superconductors performed
by Nikoli\'c and Sachdev.~\cite{Nikolic+Sachdev:06} They have
found that the effect of quasiparticles on the effective
vortex-vortex interaction (and, for that matter, some other
properties of vortices) is not very dramatic. This can probably be
ascribed to the nodal character of quasiparticle spectrum in
d-wave superconductors, as compared to the fully-gapless situation
that we are concerned with in the present work.

\subsection{Experimental parameters and conditions}

In order to elucidate the realm of validity of our effective
theory and make contact with experiments, it is useful to estimate
the physical healing (coherence) length and compare it with the
inverse of the momentum scale $k_{\Delta}$. In this section, we
discuss different regimes where our effective theory applies or
may not be relevant.

To that end, we analyze the bosonic sector of the theory. It
is known from the BEC studies~\cite{pethickbook} that the healing
length can be expressed as $\xi=(8\pi n a)^{-1/2}$, where $n$ is
bosonic density and $a$ the corresponding scattering length.
Generically, this is the length scale set by the chemical
potential of bosons ($\hbar^{2}/(2m\xi^{2})=\mu_{B}$), expressed
to lowest order in $\sqrt{na^{3}}$. Therefore, in our case the
healing length can be expressed as $\xi=(8\pi n_{s}
a_{m})^{-1/2}$, where $a_{m}$ is the molecular scattering length
and the superfluid density is the density of bosonic Feshbach
molecules.
Starting from expressions for $\xi$ and $k_{\Delta}$ we obtain
\begin{equation}\label{}
\frac{\xi}{k_{\Delta}^{-1}}=\frac{\sqrt{2m\Delta}} {\sqrt{8\pi
n_{s}a_{m}}}=\frac{\sqrt{2m\epsilon_{F}(\Delta/\epsilon_{F})}}
{\sqrt{8\pi na_{f} (n_{s}/n) (a_{m}/a_{f})}} \:\:,
\end{equation}
\noindent that is,
\begin{equation}\label{ratio}
\frac{\xi}{k_{\Delta}^{-1}}=\frac{1}{\sqrt{8\pi}}
\frac{\left(\frac{\Delta}{\epsilon_{F}}\right)^{1/2}}
{\left(\frac{n_{s}}{n}\right)^{1/2}
\left(\frac{a_{m}}{a_{f}}\right)^{1/2}}
\frac{(3\pi^{2}n)^{1/3}}{\sqrt{na_{f}}} \:\:,
\end{equation}
\noindent with $\epsilon_{F}=k_{F}^{2}/(2m)$, where
$k_{F}=(3\pi^{2}n)^{1/3}$ is the momentum scale set by the total
fermion density. The last equation can be conveniently recast as
\begin{equation}\label{brz}
\frac{\xi}{k_{\Delta}^{-1}}=\frac{(3\pi^{2})^{1/3}}{\sqrt{8\pi}}
\frac{\left(\frac{\Delta}{\epsilon_{F}}\right)^{1/2}}
{\left(\frac{n_{s}}{n}\right)^{1/2}\left(\frac{a_{m}}{a_{f}}
\right)^{1/2}}|\kappa|^{1/6} \:\:,
\end{equation}
\noindent where
\begin{equation}\label{}
 \kappa \equiv -\frac{1}{na_{f}^{3}}
\end{equation}
\noindent is a dimensionless diluteness
parameter.~\cite{Son-Stephanov:06} [Recall the  familiar results
in three important limits: $\kappa\rightarrow -\infty\:(+\infty)$
in the BEC (BCS) limit and $\kappa=0$ at unitarity.] Eq.
\eqref{brz} is equivalent to
\begin{equation}\label{xikd}
\frac{\xi}{k_{\Delta}^{-1}}=0.6171\times\frac{\left(\frac{\Delta}
{\epsilon_{F}}\right)^{1/2}}{\left(\frac{n_{s}}{n}\right)^{1/2}
\left(\frac{a_{m}}{a_{f}}\right)^{1/2}}|\kappa|^{1/6}
\end{equation}
\noindent and implies that $\xi/k_{\Delta}^{-1}$ depends on three
dimensionless ratios and the diluteness parameter.

To provide a quantitative estimate of the ratio
$\xi/k_{\Delta}^{-1}$ in the parameter regime relevant for
realization of the BP1 state, it is useful to recall the relevant
details of the mean-field phase-diagram of a polarized Fermi gas,
based on the two-channel (i.e., boson-fermion)
model.~\cite{Sheehy:06,Sheehy+Radzihovsky:07} This mean-field
theory yields quantitatively reliable results in the
narrow-resonance regime, being exact in the limit of a vanishing
resonance-width.~\cite{Andreev:04} According to this phase
diagram, for intermediate negative Feshbach-resonance detuning
($\nu$) BP1 (SF$_{\textrm{M}}$) exists in the region between lines
$\delta_{m}\approx |\nu|$ (boundary to unpolarized BEC superfluid)
and $\delta_{c1}\approx 1.3|\nu|$ (the boundary to a phase
separated state -- the superfluid-normal coexistence region).
[Note the following difference in notation: here, the
  chemical potential difference is denoted as $\del$, whereas in
Ref.~\cite{Sheehy:06,SheehyAnnals:07} it is $2h$ while the detuning
is denoted as 
$\delta$.]

It is known that in the strong-coupling BEC regime of a superfluid
Fermi gas with equal populations of two hyperfine spin components
(balanced Fermi mixtures) the molecular scattering length is given
by $a_{m}=0.6\:a_{f}$ ($a_{f}$ being the scattering length between
fermionic atoms).~\cite{Petrov+:04} For a polarized Fermi gas,
however, as shown by Sheehy and Radzihovsky (see Eq. 16 and Fig. 2
of Ref.~\cite{Sheehy+Radzihovsky:07}), the molecular scattering
length decreases monotonously as a function of $\delta/|\nu|$ (or,
equivalently, of the polarization) and vanishes at the
aforementioned boundary of first-order phase transition to the
phase separated state. Therefore, as follows from Eq.
\eqref{xikd}, right at the boundary to phase separation and in the
immediate vicinity of it the coherence length becomes much greater
than $k_{\Delta}^{-1}$, thus making the quantitative implications
of our theory not directly applicable in this special case.

Taking $a_{m}(\delta=0)$ in place of $a_{m}$, together with
typical values of $n_{s}/n$ and $\Delta/\epsilon_{F}$
($\epsilon_{F}=k_{F}^{2}/(2m)$, where $k_{F}=(3\pi^{2}n)^{1/3}$ is
the momentum scale set by the total fermion density) in the BEC
regime, we estimate that $\xi$ is of the same order as
$k_{\Delta}^{-1}$ when $|\kappa|\sim 1$-$100$. From numerical
results for $a_{m}(|\nu|,\delta)/a_{m}(|\nu|,\delta=0)$ obtained
in Ref.~\cite{Sheehy+Radzihovsky:07}, on account of the fact that
$\xi/k_{\Delta}^{-1}\propto (a_{m}/a_{f})^{-1/2}$, we can infer
that the above estimate is just slightly modified as a result of
$a_{m}$ decreasing as a function of $\delta/|\nu|$: for example,
for $\delta/|\nu|=1.2$ the true molecular scattering length is an
order of magnitude smaller than that of the unpolarized system,
but the ratio $\xi/k_{\Delta}^{-1}$ is modified only by a factor
of $\sqrt{10}\approx 3.16$. For smaller values of $\delta/|\nu|$
this factor is even smaller, i.e., it is of the order of unity.
Thus, this estimate confirms that our choice of $k_{\Delta}$ as
the upper momentum cutoff of the theory is physically pertinent.

Moreover, using the expression of $\Delta/\epsilon_{F}$ in
Ref.~\cite{engelbrecht:97}, we can straightforwardly infer that in
the BEC limit $\xi/k_{\Delta}^{-1}\propto |\kappa|^{1/4}$. It
follows that $\xi/k_{\Delta}^{-1}\rightarrow \infty$ in the BEC
limit, which seems to suggest that this limit is out of the
application scope of our theory, since the latter is intrinsically
valid for physics at distances longer than $\xi$.  It is, however,
important to emphasize that the effect of gapless fermions on the
interaction between vortices is not even expected to bear any
physical relevance in the BEC limit, where the system at hand
essentially becomes a Bose-Fermi mixture akin to the $^3$He-$^4$He
mixture.~\cite{nishida+son:06} Technically speaking, this point is
manifest in the Nishida-Son formulation of the effective
Lagrangian for the imbalanced Fermi gas, through a vanishing
coupling coefficient between the fermion current and the gradient
of phase field (supercurrent) in this limit. In our case,
largeness of the physically-allowed inter-vortex distance scale
($r\gtrsim \xi$) compared to $k_{\Delta}^{-1}$ in the BEC limit
and the fact that at very long distances (compared to
$k_{\Delta}^{-1}$) the fermion-induced part of the vortex-vortex
interaction is quantitatively unimportant compared to the
conventional (infinitely-ranged) repulsive logarithmic
contribution are indeed suggesting that the physical effect under
consideration is absent in this limit. This is an important
consistency check of our results.

\section{Summary and conclusions}

In summary, starting from a Lagrangian for the superfluid phase
field and the gapless branch of fermionic quasiparticles, we have
obtained the effective action for vortices in a spin-polarized
homogeneous superfluid state with a single gapless Fermi surface.
We have demonstrated that besides the conventional repulsive
logarithmic part (2D Coulomb potential) the effective vortex
interaction potential has an additional, predominantly attractive,
component induced by the presence of gapless fermions. This
fermion-induced potential has oscillating character analogous to
the RKKY indirect-exchange interaction. Interactions between
defects mediated by the continuum they are immersed in (either
bosonic or fermionic) have been studied quite recently in several
different physical contexts and different
dimensionalities.~\cite{Recati:05,Fuchs:07} Our work, however,
constitutes the first study of this kind that concerns the
interaction between vortices in superfluids. It shows that besides
the Friedel oscillations (charge sector) and the RKKY (spin
sector), an analogous oscillating phenomenon appears in the vortex
sector.

Our study opens up a question as to the nature of the vortex
lattice in gapless fermionic superfluids. Due to the partly
attractive nature of the effective vortex potential that we have
found, the resulting vortex lattice structure in BP1 superfluid
phase could be different than the triangular lattice, which would be a
spectacular experimental signature. The complexity of the problem,
however, calls for an elaborate future study. Even when the
potential has a unique distance dependence of a known analytical
form (such as, for example, the conventional logarithmic
interaction), calculation of the resulting lattice structure is
quite a nontrivial task, since
lattice-summation-methods~\cite{campbelldoria:89,longrangesums}
for long-range potentials are strongly dependent on the actual
form of the potential. The new vortex-vortex interaction potential
is not obtained, due to the complexity of the problem, in a closed
analytical form and has both attractive and repulsive parts. This
unusual, non-monotonous, distance dependence of the effective
potential implies that vortex lattice structure may in fact not be
unique, but also depend on the geometrical constraints on the
system, for example, the range of distances between individual
vortices realized for a given size of the superfluid container.
The standard lattice summation methods may not be applicable and
more sophisticated strategies need to be employed, for instance
Monte Carlo calculations.

In the present work we have studied the intrinsic effect of
gapless fermionic excitations on the interaction between vortices
in the BP1 state and have therefore considered only the
homogeneous case. Our results are expected to be also valid for a
trapped system as long as the trap potential varies smoothly on
the scale of the Fermi wavelength (or, more generally, the longest
physical length-scale in the problem), that is, in the regime of
validity of the local density approximation. However, an important
problem yet to be explored is the possible influence of strong
spatial inhomogeneities caused by the presence of the trap on the
form of the vortex lattice, as studied by Sheehy and Radzihovsky
in the context of trapped Bose
gases.~\cite{Sheehyvort1:04,Sheehyvort2:04} They have provided an
explanation for the striking uniformity of the vortex lattices
seen in experiments in spite of the strong spatial variation of
the local superfluid density imposed by the trap. Moreover, they
have shown that an interplay of an inhomogeneous trap potential
and vortex discreteness leads to a vortex density that is largest
in the center of the trap, a counterintuitive result from the
energetic point of view because both the kinetic energy cost and
the repulsive interaction between vortices are proportional to the
local superfluid density and are therefore largest in the center
of the trap. As we have shown in the present study of a
spin-polarized Fermi gas, for sufficiently high polarization the
effective interaction between vortices in this system is
attractive at short distances and could therefore bring about some
completely new effects, such as the competition between this
attractive interaction and the kinetic energy cost. Further
investigation of the properties of ``vortex matter'' in
spin-polarized Fermi gases is thus clearly called for.

\section*{Acknowledgments}

We thank D. T. Son for useful discussions. V.M.S. and W.V.L. were
supported in part by the ORAU and ARO (W911NF-07-1-0293). Y.B.K.
was supported by the NSERC, CRC, CIAR, KRF-2005-070-C00044, and
the Miller Institute for Basic Research in Science at UC Berkeley
through the Visiting Miller Professorship. This research was
supported during the completion in part in KITP at UCSB by the
National Science Foundation under Grant No. NSF PHY05-51164.

\appendix
\section{Galilean invariance}  \label{Galilean}
In this Appendix, we explicitly demonstrate the Galilean invariance
of the fermion-dependent part of Lagrangian \eqref{lagr}. We shall
first establish an explicit relation of the quasiparticle field in the present
effective field theory to the fermion particle field in a microscopic model,
and then derive the Galilean transformation properties of the quasiparticle
field from that of the (microscopic) fermion fields. Subsequently, an
alternative approach will be given to provide a further
justification and understanding.

\subsection{Microscopic relation of the quasiparticle field}

To examine the Galilean transformation of the Bogoliubov
quasiparticle field, let us consider a microscopic model of
Lagrangian \eqref{micrlagr}. As a result of the
Hubbard-Stratonovich transformation in the Cooper channel,
introducing the auxiliary pair field $\Delta(x)$, this Lagrangian
changes to
\begin{equation}
\tilde{\mathcal{L}}=\psi_{\sigma}^{*}
\left(\partial_{\tau}-\frac{\nabla^{2}}{2m_{\sigma}}
-\mu_{\sigma}\right)\psi_{\sigma}+(\psi_{\uparrow}^{*}
\psi_{\downarrow}^{*}\Delta(x)+\textrm{c.c})
+\frac{1}{g}|\Delta(x)|^{2}  \:\:.
\end{equation}
\noindent [Summation over repeated pseudo-spin indices in the last
equation is implicit.] Ignoring fluctuations of the amplitude of
the order parameter, i.e., assuming that
$\Delta(x)=\Delta\:e^{i\theta(x)}$, it is advantageous to
transform the fermion fields at each space-time point
as~\cite{liu:eft:06}
\begin{equation}
\psi_{\sigma}(x)=\tilde{\psi}_{\sigma}(x)
\:e^{\frac{i}{2}\theta(x)} \:\:\:, \qquad
\psi^{*}_{\sigma}(x)=\tilde{\psi}^{*}_{\sigma}(x)
\:e^{-\frac{i}{2}\theta(x)} \:\:.
\end{equation}
\noindent This local (gauge) transformation is designed to
transform away the phase-fluctuation dependence from the
off-diagonal pairing potential terms to the diagonal (kinematic)
terms in the fermion sector of the theory. As a result, the
$\tilde{\psi}_{\sigma}$ fermion fields are locally stripped off of
any dependence on the U(1) phase $\theta(x)$. The transformed
Lagrangian can be written as $\tilde{\mathcal{L}}=
\tilde{\mathcal{L}}_{0}+\mathcal{L}_{\tilde{\psi},\theta}$, where
\begin{equation}\label{}
\tilde{\mathcal{L}}_{0}= \tilde{\psi}_{\sigma}^{*}
\left(\partial_{\tau}-\frac{\nabla^{2}}{2m_{\sigma}}
-\mu_{\sigma}\right)\tilde{\psi}_{\sigma}+
(\Delta\tilde{\psi}_{\uparrow}^{*}\tilde{\psi}_{\downarrow}^{*}
+\textrm{c.c})
\end{equation}
\noindent is the mean-field Lagrangian for $\tilde{\psi}_{\sigma}$
fermions, and
\begin{equation}\label{}
\mathcal{L}_{\tilde{\psi},\theta}=
\tilde{\psi}_{\sigma}^{*}\tilde{\psi}_{\sigma}
\left(i\partial_{\tau}\theta+\frac{1}{2m_{\sigma}}(\nabla\theta)^{2}
\right)-\frac{i}{2m_{\sigma}}\left(\tilde{\psi}_{\sigma}^{*}
\nabla\tilde{\psi}_{\sigma}-\nabla\tilde{\psi}_{\sigma}^{*}
\tilde{\psi}_{\sigma}\right)\cdot\nabla\theta \:\:.
\end{equation}
\noindent With $\tilde{\mathcal{L}}_{0}$ naturally giving rise to
the Bogoliubov quasiparticles as its elementary excitations,
$\mathcal{L}_{\tilde{\psi},\theta}$ essentially contains, in an
implicit form, all the couplings of these excitations to the
superfluid phase fluctuations.

A cautious remark is needed for the Lagrangian derived above.  It
appears that we have just provided a derivation for the postulated
effective Lagrangian \eqref{lagr}. One may be tempted to determine
the ``phenomenological'' coefficients of this Lagrangian in this
way. For weak coupling, this can indeed be done. For a strongly
interacting Fermi gas, the derivation from the microscopic model
cannot be done in a controlled approximation, once the pairing
amplitude and density fluctuations are included. The
symmetry-based Lagrangian of the postulated form \eqref{lagr}
describes the same physics, albeit from a more phenomenological
point of view. Moreover, it does not suffer from the difficulty in
strong coupling. In summary, the above derivation is understood to
provide an example of how to separate the low energy Goldstone
bosons (the phase fluctuation) from other degrees of freedom, but
not a rigorous proof of the effective Lagrangian itself on a
microscopic level.

\subsection{Galilean transformation for quasiparticles}

Let us denote the laboratory frame as $K$ and the corresponding
spatial and time coordinates as $\mathbf{x}$ and $t$. We shall
also denote a frame moving with velocity $\mathbf{u}$ relative to
$K$ as $K'$, and its spatial and time coordinates as $\mathbf{x'}$
and $t'$. Under Galilean boost transformation with velocity
$\mathbf{u}$ the space-time coordinates transform as :
$\mathbf{x}\rightarrow \mathbf{x'}=\mathbf{x}-\mathbf{u}t$,
$t\rightarrow t'=t$. The spatial and time derivatives transform as
\begin{equation}\label{derprop}
\nabla\longrightarrow\nabla'=\nabla \:\:,\qquad
\partial_{\tau}\longrightarrow \partial'_{\tau'}=
\partial_{\tau}-i(\mathbf{u}\cdot\nabla)\:\:.
\end{equation}

Under the above Galilean transformation, the (microscopic) fermion
field of mass $m$ transforms in the standard way, \be \psi(x)
\longrightarrow \psi^\prime(x^\prime)= e^{i(-m\mathbf{u}\cdot \Bx+
{1\over 2}m \mathbf{u}^2 t)} \psi(x) \ee where $x\equiv
(\Bx,\tau=it)$. Being locally stripped off of any `charge'
U$_c$(1) phase dependence, the $\tilde{\psi}_{\sigma}$ fermion
fields are by construction invariant under Galilean
transformation: \be \tilde{\psi}_{\sigma}(x) \longrightarrow
\tilde{\psi}_{\sigma}'(x')=\tilde{\psi}_{\sigma}(x)\,,
\label{eq:tpsi:GT} \ee and the `charge' U$_c$(1) phase transforms
\begin{equation}\label{thetaprop}
\theta(x)\longrightarrow \theta'(x')=
\theta(x)-2m\mathbf{u}\cdot\mathbf{x}
-im\mathbf{u}^{2}\tau \:\:.
\end{equation}
The
Bogoliubov quasiparticles (the two branches being denoted by
$\chi_{\uparrow}$ and $\chi_{\downarrow}$) can now be introduced
as
\begin{eqnarray}
\chi_{\uparrow}(\mathbf{k},\tau) &=&  u_{\mathbf{k}}
\tilde{\psi}_{\uparrow}(\mathbf{k},\tau)
+v_{\mathbf{k}}\tilde{\psi}^{*}_{\downarrow}(-\mathbf{k},\tau) \:\:,\\
\chi_{\downarrow}(\mathbf{k},\tau) &=& u_{-\mathbf{k}}
\tilde{\psi}_{\downarrow}(\mathbf{k},\tau)-v_{-\mathbf{k}}
\tilde{\psi}^{*}_{\uparrow}(-\mathbf{k},\tau) \:\:,
\end{eqnarray}
\noindent where $u_{\mathbf{k}},v_{\mathbf{k}}$ are Bogoliubov
amplitudes. For the lower branch quasiparticle field, which is
simply denoted by $\chi$ (i.e, $\chi\equiv \chi_\down$), this
becomes in real space
\begin{equation}\label{}
\chi (\mathbf{x},\tau)=\int
d\mathbf{y}\left[u(\mathbf{y-x})
\tilde{\psi}_{\downarrow}(\mathbf{y},\tau)-v(\mathbf{x-y})
\tilde{\psi}^{*}_{\uparrow}(\mathbf{y},\tau)\right]
\end{equation}
By using \eqref{eq:tpsi:GT}, we find
\be
\chi(x) \longrightarrow \chi^\prime(x^\prime) =\chi(x)\,. \label{eq:chi'=}
\ee
From the transformation properties of $\chi$ and $\theta$, it is
straightforward to prove the
Galilean-invariance of the term
\begin{equation}\label{}
\chi^{*}\chi\left\{i\partial_{\tau}
\theta+\frac{1}{2m_{p}}(\nabla\theta)^{2}\right\}=\chi^{*}\chi
U_{\theta}
\end{equation}
\noindent in Lagrangian \eqref{lagr}. Besides, using relations
\eqref{thetaprop} and \eqref{derprop} (the latter implies that
$\varepsilon(-i\nabla)$ is an invariant), we can readily prove
that the combination
\begin{equation}\label{}
\chi^{*}[\partial_{\tau}+\varepsilon(-i\nabla)]\chi+
\nabla\theta\cdot\mathbf{j}
\end{equation}
\noindent is Galilean invariant up to an unimportant total
derivative. This proves the Galilean invariance
for the fermion-dependent part of Lagrangian \eqref{lagr}.

\subsection{The Doppler shift}

An alternative check of the Galilean invariance of the
quasiparticle field can be obtained by starting from the
requirement that the quasiparticle energy is Doppler shifted under
a Galilean boost.

We first review a standard derivation of the Galilean transformation
~\cite{Landau-Lifshitz:77QM}. Recall how momentum and energy of particles with
quadratic dispersion (e.g., bare fermions) transform under this
Galilean boost :
\begin{equation}\label{}
\mathbf{p}\longrightarrow \mathbf{p}'=\mathbf{p}-m\mathbf{u}
\:\:,\qquad E  \longrightarrow
E^\prime=E -\mathbf{p}\cdot\mathbf{u}
+\frac{1}{2}m\mathbf{u}^{2} \:\:.
\end{equation}
\noindent Using these rules, it is straightforward to show that
the combination $\mathbf{p}\cdot\mathbf{x}-E  t$ shifts
by a factor
$-m\mathbf{u}\cdot\mathbf{x}+\frac{1}{2}m\mathbf{u}^{2}t$, which
depends on the parameters of the transformation ($m,\mathbf{u}$)
but does not depend on $\mathbf{p}$. Accordingly, every plane wave
\begin{equation}\label{plwave}
 \varphi_{\mathbf{p}}(\mathbf{x},t)=\mathrm{const}\times
 e^{i(\mathbf{p}\cdot\mathbf{x}-E  t)}
\end{equation}
\noindent acquires the same phase factor
$\exp[i(-m\mathbf{u}\cdot\mathbf{x}+\frac{1}{2}m\mathbf{u}^{2}t)]$
under the Galilean boost, regardless of $\mathbf{p}$. Moreover,
since an arbitrary single particle wave-function can be expanded
in plane-waves \eqref{plwave}, we conclude that each wave function
picks up that same phase factor under this boost. Because Galilean
transformations are space-time symmetry transformations, the
transformation property of the single-particle wave-function
carries over to the field operators
\begin{equation}\label{}
\hat{\psi}(\mathbf{x},t)=\sum_{n}
\hat{a}_{n}\phi_{n}(\mathbf{x},t)
\:\:,
\end{equation}
\noindent where $\phi_{n}(\mathbf{x},t)$ form an arbitrary
complete orthonormal set of single particle states.

Bearing in mind the definition \eqref{phasedef}, as a by-product of
the transformation rule found above, we conclude that the
superfluid phase field is transformed
as~\cite{Landau-Lifshitz:77QM}
\begin{equation}\label{}
\theta(\mathbf{x},t)\longrightarrow\theta'(\mathbf{x'},t')=
\theta(\mathbf{x},t)-m_{p}\mathbf{u}\cdot\mathbf{x}
+\frac{1}{2}m_{p}\mathbf{u}^{2}t \:\:,
\end{equation}
\noindent that is, we recover transformation \eqref{thetaprop} when
changing over to the imaginary time.
\noindent By making use of the transformation properties
\ref{derprop} it is straightforward to show that the combination
\begin{equation}\label{}
U_{\theta}=i\partial_{\tau}\theta+\frac{1}{2m_{p}}(\nabla\theta)^{2}
\end{equation}
\noindent is invariant under the Galilean transformation.

In order to determine how the Bogoliubov quasiparticle field
transforms under the Galilean transformation, we recall that the
momentum of a quasiparticle is invariant under the Galilean
transformation while the quasiparticle energy is Doppler-shifted
(to leading order in the boost velocity):~\cite{VolovikReports:01}
\begin{equation}\label{}
\mathbf{p}\longrightarrow \mathbf{p}'=\mathbf{p} \:\:,\qquad
E  \longrightarrow
E'=E -\mathbf{p}\cdot\mathbf{u} \:\:.
\end{equation}
\noindent Based on these properties, it is easy to demonstrate
that for Bogoliubov quasiparticles the combination
$\mathbf{p}\cdot\mathbf{x}-Et$ remains invariant under the
Galilean transformations (independent of $\mathbf{p}$ and $E $),
which using analogous reasoning as above implies that an arbitrary
single-particle wave-function and the field operator
$\hat{\chi}(\mathbf{x},t)$ of a Bogoliubov quasiparticle is
invariant under Galilean transformations :
\begin{equation}\label{}
\hat{\chi}(\mathbf{x},t) \longrightarrow
\hat{\chi'}(\mathbf{x'},t')=\hat{\chi}(\mathbf{x},t) \:\:.
\end{equation}
\noindent This is the equivalent form of \eqref{eq:chi'=} in
operator formalism.

\section{Expression for the transverse current response function}\label{pqo}

Because $\varepsilon_{\mathbf{k}}>0$ for $|\mathbf{k}|>k_{b}$ and
$n_{F}(\varepsilon)\rightarrow \theta(-\varepsilon)$ as
$T\rightarrow 0$, in the zero-temperature static limit the
response function $P(q)$ (defined by Eq.~\eqref{pqdef}) reduces to
\begin{equation}\label{}
P^{0}_{\mathbf{q}}
=\frac{1}{2}\int\frac{d^{3}\mathbf{k}}{(2\pi)^{3}}
\left[\frac{\theta(|\mathbf{k+q}|-k_{b})
\theta(k_{b}-|\mathbf{k}|)}{\varepsilon_{\mathbf{k}}-
\varepsilon_{\mathbf{k+q}}+i\eta}-\frac{\theta(k_{b}-
|\mathbf{k+q}|)
\theta(|\mathbf{k}|-k_{b})}{\varepsilon_{\mathbf{k}}-
\varepsilon_{\mathbf{k+q}}-i\eta}\right]\mathbf{k}_{\perp}^{2}\:\:,
\end{equation}
\noindent where $\eta\rightarrow 0+$ and the momentum sum in
\eqref{pqdef} has been replaced by an integral. We now undertake
the change of variables $\mathbf{k}'=-\mathbf{k}-\mathbf{q}$ in
the second term of the last equation. Because
$\mathbf{k'}_{\perp}=-\mathbf{k}_{\perp}$, we have that
$\mathbf{k'}_{\perp}^{2}= \mathbf{k}_{\perp}^{2}$. In other words,
$\mathbf{k}_{\perp}^{2}$ is invariant under this change of
variables. Consequently, we arrive at
\begin{equation}\label{}
P^{0}_{\mathbf{q}}=\frac{1}{2}\int
\frac{d^{3}\mathbf{k}}{(2\pi)^{3}}
\:\theta(|\mathbf{k+q}|-k_{b})\theta(k_{b}-
|\mathbf{k}|)\left[\frac{1}{\varepsilon_{\mathbf{k}}-
\varepsilon_{\mathbf{k+q}}+i\eta}
-\frac{1}{\varepsilon_{\mathbf{k+q}}-
\varepsilon_{\mathbf{k}}-i\eta}\right] \mathbf{k}_{\perp}^{2}\:\:,
\end{equation}
\noindent where the superfluous prime has been omitted (i.e., we
have returned to the initial integration variable $\mathbf{k}$).
The last equation can obviously be simplified to
\begin{equation}\label{}
P^{0}_{\mathbf{q}}=\int\frac{d^{3}\mathbf{k}}{(2\pi)^{3}}
\:\theta(|\mathbf{k+q}|-k_{b})
\theta(k_{b}-|\mathbf{k}|)\:\frac{\mathbf{k}_{\perp}^{2}}
{\varepsilon_{\mathbf{k}}-\varepsilon_{\mathbf{k+q}}+i\eta}\qquad
(\: \eta \rightarrow 0+\:) \:\:\:.
\end{equation}
\noindent By virtue of the Sohotsky-Plemelj formula
\begin{equation}\label{sohplem}
 \lim_{\eta\searrow 0}\frac{1}{x\pm i\eta}=\mathcal{P}
 \frac{1}{x}\mp i\pi\delta(x)\:\:,
\end{equation}
\noindent we can now demonstrate that $\mathrm{Im}\:
\{P^{0}_{\mathbf{q}}\}=0$ and that $\mathrm{Re}\:
\{P^{0}_{\mathbf{q}}\}=P^{0}_{\mathbf{q}}$ is given by
\begin{equation}\label{}
P^{0}_{\mathbf{q}}=\mathcal{P}\int
\frac{d^{3}\mathbf{k}}{(2\pi)^{3}}\:\theta(|\mathbf{k+q}|-k_{b})
\theta(k_{b}-|\mathbf{k}|)\:\frac{\mathbf{k}_{\perp}^{2}}
{\varepsilon_{\mathbf{k}}-\varepsilon_{\mathbf{k+q}}}\:\:,
\end{equation}
\noindent where $\mathcal{P}$ stands for the Cauchy principal
value. Using the identity $\theta(x)=1-\theta(-x)$ for
$x=|\mathbf{k+q}|-k_{b}$, the last equation becomes
\begin{equation}\label{}
P^{0}_{\mathbf{q}}=
\mathcal{P}\int\frac{d^{3}\mathbf{k}}{(2\pi)^{3}}
\Big[1-\theta(k_{b}-|\mathbf{k+q}|)\Big]
\theta(k_{b}-|\mathbf{k}|)\:\frac{\mathbf{k}_{\perp}^{2}}
{\varepsilon_{\mathbf{k}}-\varepsilon_{\mathbf{k+q}}}\:\:.
\end{equation}
\noindent The term that contains the product of two step functions
vanishes identically after the integration, because this product
is even under the interchange $\mathbf{k}\leftrightarrows
\mathbf{k+q}$, while the fraction
$\mathbf{k}_{\perp}^{2}/(\varepsilon_{\mathbf{k}}
-\varepsilon_{\mathbf{k+q}})$ is odd under the same transformation
(while $\varepsilon_{\mathbf{k}}-\varepsilon_{\mathbf{k+q}}$ is
obviously odd, the fact that
$\mathbf{k}_{\perp}=(\mathbf{k+q})_{\perp}$ implies that
$\mathbf{k}_{\perp}^{2}$ is even) ; accordingly, we have
\begin{equation}\label{}
P^{0}_{\mathbf{q}}=\mathcal{P}\int
\frac{d^{3}\mathbf{k}}{(2\pi)^{3}}
\:\theta(k_{b}-|\mathbf{k}|)\:\frac{\mathbf{k}_{\perp}^{2}}
{\varepsilon_{\mathbf{k}}-\varepsilon_{\mathbf{k+q}}}\:\:.
\end{equation}

With the aid of identity
$\mathbf{k}_{\perp}^{2}=|\mathbf{k}|^{2}(1-\cos^{2}\theta)$ and
momentum re-scaling $\mathbf{k}/k_{b}\rightarrow\mathbf{k}$ (such
that all momenta are expressed in units of $k_{b}$) we express
$P^{0}_{\mathbf{q}}$ as a principal-value integral over the
dimensionless momentum :
\begin{equation}\label{}
P^{0}_{\mathbf{q}}=\frac{k_{b}^{3}}{(2\pi)^{2}}\mathcal{P}
\int_{0}^{1}|\mathbf{k}|^{4}d|\mathbf{k}|
\int_{0}^{\pi}\frac{1-\cos^{2}\theta}{\varepsilon_{\mathbf{k}}-
\varepsilon_{\mathbf{k+q}}}\sin\theta\:d\theta  \:\:.
\end{equation}
\noindent Finally, upon inserting dispersion~\eqref{disp} and
making substitution $x=\cos\theta$, this integral leads to
\begin{equation}\label{}
P^{0}_{\mathbf{q}}=\frac{k_{b}^{3}}{(2\pi)^{2}}\mathcal{P}
\int_{0}^{1}|\mathbf{k}|^{4}d|\mathbf{k}|\int_{-1}^{1}dx
\frac{1-x^{2}}{\sqrt{\xi_{\mathbf{k}}^{2}+\left(\frac{\Delta}{k_{b}^2}
\right)^{2}}-\sqrt{\left(\xi_{\mathbf{k}}+\frac{|\mathbf{q}|^2}{2m}
+\frac{|\mathbf{k}||\mathbf{q}|}{m}x\right)^2+
\left(\frac{\Delta}{k_{b}^2}\right)^{2}}}\:\:,
\end{equation}
\noindent where $\xi_{\mathbf{k}}\equiv |\mathbf{k}|^{2}/2m-
(\mu/k_{b}^{2})$.

\section{Behavior of $P^{0}_{\mathbf{q}}$ for $|\mathbf{q}|\ll k_{b}$
and the $|\mathbf{q}|\rightarrow 0$ limit} \label{anres}

The most general expression for $P(q)\equiv
P(\mathbf{q},i\omega_{l})$ reads
\begin{equation}\label{resp1}
P(\mathbf{q},i\omega_{l})=\frac{1}{2V} \sum_{\mathbf{k}}
\frac{n_{F}(\varepsilon_{\mathbf{k}})
-n_{F}(\varepsilon_{\mathbf{k+q}})}
{i\omega_{l}+\varepsilon_{\mathbf{k}} -\varepsilon_{\mathbf{k+q}}}
\:\mathbf{k}_{\perp}^{2}\:\:,
\end{equation}
\noindent that is
\begin{equation}\label{resp2}
P(\mathbf{q},i\omega_{l})=\frac{1}{2}\int
\frac{d^{3}\mathbf{k}}{(2\pi)^{3}}
\frac{n_{F}(\varepsilon_{\mathbf{k}})
-n_{F}(\varepsilon_{\mathbf{k+q}})}
{i\omega_{l}+\varepsilon_{\mathbf{k}} -\varepsilon_{\mathbf{k+q}}}
\:\mathbf{k}_{\perp}^{2}\:\:.
\end{equation}

In order to calculate $P^{0}_{\mathbf{q}}$ for $|\mathbf{q}|\ll
k_{b}$, we start from the expansion
\begin{equation}\label{expnf}
n_{F}(\varepsilon_{\mathbf{k}})-n_{F}(\varepsilon_{\mathbf{k+q}})=
\frac{\partial n_{F} (\varepsilon_{\mathbf{k}})}{\partial
\varepsilon_{\mathbf{k}}}\:(\varepsilon_{\mathbf{k}}-
\varepsilon_{\mathbf{k+q}})+\mathcal{O}(|\mathbf{q}|^{2}) \:\:,
\end{equation}
\noindent valid for $|\mathbf{q}|\ll k_{b}$. At zero temperature
$n_{F}(\varepsilon)=\theta(-\varepsilon)$, implying that $\partial
n_{F}(\varepsilon)/\partial\varepsilon=-\delta(\varepsilon)$. For
linearized dispersion $\varepsilon_{\mathbf
k}=v_{b}(|\mathbf{k}|-k_{b})$, using the fact that
$\delta(cx)=\delta(x)/|c|$, we find
\begin{equation}\label{nf}
n_{F}(\varepsilon_{\mathbf{k}})-n_{F}(\varepsilon_{\mathbf{k+q}})=
(|\mathbf{k+q}|-|\mathbf{k}|)\:\delta(|\mathbf{k}|-k_{b})
+\mathcal{O}(|\mathbf{q}|^{2})\:\:.
\end{equation}
\noindent Here $\mathbf{k}\cdot\mathbf{q}=|\mathbf{k}||\mathbf{q}|
\cos\theta$, and consequently
$|\mathbf{k+q}|=(|\mathbf{k}|^{2}+2|\mathbf{k}|
|\mathbf{q}|\cos\theta+|\mathbf{q}|^{2})^{1/2}$.

On account of result \eqref{nf}, together with
$\mathbf{k}_{\perp}^{2}=|\mathbf{k}|^{2}(1-\cos^{2}\theta)$, Eq.
\eqref{resp2} leads to an integral (trivial integration over the
azimuthal angle yields factor $2\pi$)
\begin{equation}\label{}
P(\mathbf{q},i\omega_{l})\simeq \frac{1}{2(2\pi)^{2}}
\int_{0}^{\infty}|\mathbf{k}|^{4}d|\mathbf{k}|
\int_{-1}^{1}d(\cos\theta)\frac{(1-\cos^{2}\theta)
(|\mathbf{k+q}|-|\mathbf{k}|)
\delta(|\mathbf{k}|-k_{b})}{i\omega_{l}
-v_{b}(|\mathbf{k+q}|-|\mathbf{k}|)}\:\:.
\end{equation}
\noindent Upon executing the integral over $|\mathbf{k}|$ and
introducing substitution $x=\cos\theta$, we arrive at
\begin{equation}\label{kint}
P(\mathbf{q},i\omega_{l})\simeq \frac{k_{b}^{4}}{2(2\pi)^{2}}
\int_{-1}^{1}\frac{(1-x^{2})\left(\sqrt{k_{b}^{2}+
2k_{b}|\mathbf{q}|x+|\mathbf{q}|^{2}}-k_{b}\right)}{i\omega_{l}
-v_{b}\left(\sqrt{k_{b}^{2}+
2k_{b}|\mathbf{q}|x+|\mathbf{q}|^{2}}-k_{b}\right)}\:dx \:\:.
\end{equation}
\noindent Another variable substitution
$t=\sqrt{k_{b}^{2}+2k_{b}|\mathbf{q}|x+|\mathbf{q}|^{2}}$ turns
the last integral into
\begin{equation}\label{tint}
 P(\mathbf{q},i\omega_{l})\simeq \frac{k_{b}}{8(2\pi)^{2}v_{b}
 |\mathbf{q}|^{3}}
 \int_{k_{b}-|\mathbf{q}|}^{k_{b}+|\mathbf{q}|} \frac{t(t-k_{b})
 \{(t^{2}-k_{b}^{2}-|\mathbf{q}|^{2})^{2}-(2k_{b}|\mathbf{q}|)^{2}\}}
 {t-k_{b}-i\frac{\omega_{l}}{v_{b}}}\:dt \:\:.
\end{equation}
\noindent By carrying out this integral and taking the static
limit $\omega_{l}\rightarrow 0$, we obtain the result (without the
prefactor) $-\frac{16}{3}k_{b}^{3}|\mathbf{q}|^{3}$, implying that
the first order term in expansion \eqref{expnf} yields the
$\mathbf{q}$-independent contribution
\begin{equation}\label{poq}
-\frac{k_{b}^{4}}{6\pi^{2}v_{b}}
\end{equation}
\noindent to $P^{0}_{\mathbf{q}}$.  In a similar manner,
lengthy but otherwise straightforward calculation shows that the
next (second-order) term in expansion \eqref{expnf}, namely
\begin{equation}\label{}
\frac{1}{2}\frac{\partial^{2} n_{F}
(\varepsilon_{\mathbf{k}})}{\partial
\varepsilon_{\mathbf{k}}^{2}}\:(\varepsilon_{\mathbf{k}}-
\varepsilon_{\mathbf{k+q}})^{2}=-\frac{1}{2}
(|\mathbf{k+q}|-|\mathbf{k}|)^{2} \delta'(|\mathbf{k}|-k_{b})
\:\:,
\end{equation}
\noindent adds the contribution
\begin{equation}\label{}
 -\frac{k_{b}^{2}}{10\pi^{2}v_{b}}|\mathbf{q}|^{2}
 +\frac{1}{420\pi^{2}v_{b}}|\mathbf{q}|^{4} \:\:.
\end{equation}
\noindent Therefore, for $|\mathbf{q}|\ll k_{b}$ this response
function is given by
\begin{equation}\label{}
P^{0}_{\mathbf{q}} = -\frac{k_{b}^{4}}{6\pi^{2}v_{b}}
-\frac{k_{b}^{2}}{10\pi^{2}v_{b}}|\mathbf{q}|^{2}
+\mathcal{O}(|\mathbf{q}|^{4}) \:\:,
\end{equation}
\noindent implying that
\begin{equation}\label{}
 P^{0}_{\mathbf{q}} \rightarrow -\frac{k_{b}^{4}}{6\pi^{2}v_{b}}
 \qquad (\: |\mathbf{q}|\rightarrow 0 \:)  \:\:.
\end{equation}

\section{Calculation of $R^{0}_{ij}(\mathbf{q})$ in the
$|\mathbf{q}|\rightarrow 0$ limit} \label{rqocalc}

The most general expression for $R_{ij}(q)\equiv
R_{ij}(\mathbf{q},i\omega_{l})$ reads
\begin{equation}
R_{ij}(\mathbf{q},i\omega_{l})=\frac{1}{V}
\sum_{\mathbf{k}}\frac{n_{F}(\varepsilon_{\mathbf{k}})
-n_{F}(\varepsilon_{\mathbf{k+q}})}
{i\omega_{l}+\varepsilon_{\mathbf{k}}-\varepsilon_{\mathbf{k+q}}}
\:\left(k_{i}+\frac{q_{i}}{2}\right)
\left(k_{j}+\frac{q_{j}}{2}\right) \:\:,
\end{equation}
\noindent that is
\begin{equation}
R_{ij}(\mathbf{q},i\omega_{l})=\int
\frac{d^{3}\mathbf{k}}{(2\pi)^{3}}
\frac{n_{F}(\varepsilon_{\mathbf{k}})
-n_{F}(\varepsilon_{\mathbf{k+q}})}
{i\omega_{l}+\varepsilon_{\mathbf{k}} -\varepsilon_{\mathbf{k+q}}}
\:\left(k_{i}+\frac{q_{i}}{2}\right)
\left(k_{j}+\frac{q_{j}}{2}\right) \:\:.
\end{equation}

We first show that $R_{ij}(\mathbf{q},i\omega_{l})=0$ for $i\neq
j$. To that end, we perform a rotation of the coordinate system
around the $z$-axis that maps the $x$-axis onto the $y$-axis and
the $y$-axis onto -$x$. Knowing that the module of the jacobian of
this transformation (rotation) is unity and that
$\varepsilon_{\mathbf{k}}$ depends only on $|\mathbf{k}|$ (which
is invariant under this transformation) we obtain that
$R_{xy}(q)=-R_{xy}(q)$ and $R_{yx}(q)=-R_{yx}(q)$, which implies
that $R_{xy}(q)=R_{yx}(q)=0$.

In order to calculate
\begin{equation}
R_{ii}(\mathbf{q},i\omega_{l})=\int
\frac{d^{3}\mathbf{k}}{(2\pi)^{3}}
\frac{n_{F}(\varepsilon_{\mathbf{k}})
-n_{F}(\varepsilon_{\mathbf{k+q}})}
{i\omega_{l}+\varepsilon_{\mathbf{k}} -\varepsilon_{\mathbf{k+q}}}
\:\left(k_{i}+\frac{q_{i}}{2}\right)^{2}
\end{equation}
\noindent we perform a rotation of the coordinate system that maps
the $i$-axis onto the $z$-axis, while leaving the remaining axis
invariant. $R_{ii}(\mathbf{q},i\omega_{l})$ then becomes
\begin{equation}
R_{ii}(\mathbf{q},i\omega_{l})=\int
\frac{d^{3}\mathbf{k'}}{(2\pi)^{3}}
\frac{n_{F}(\varepsilon_{\mathbf{k'}})
-n_{F}(\varepsilon_{\mathbf{k'+q}})}
{i\omega_{l}+\varepsilon_{\mathbf{k'}}
-\varepsilon_{\mathbf{k'+q}}}
\:\left(k'_{z}+\frac{q_{z}}{2}\right)^{2}
\end{equation}
\noindent for both $i=x$ and $i=y$. Thus
$R_{xx}(q)=R_{yy}(q)=R(q)$, and since the last integral can depend
only on $|\mathbf{q}|$ we can choose $\mathbf{q}$ to lie along the
$z$-axis, in which case $R(q)$ can be expressed as
\begin{equation}
R(\mathbf{q},i\omega_{l})=\int\frac{d^{3}\mathbf{k'}}{(2\pi)^{3}}
\frac{n_{F}(\varepsilon_{\mathbf{k'}})
-n_{F}(\varepsilon_{\mathbf{k'+q}})}
{i\omega_{l}+\varepsilon_{\mathbf{k'}}
-\varepsilon_{\mathbf{k'+q}}}
\:\left(k'_{z}+\frac{|\mathbf{q}|}{2}\right)^{2} \:\:,
\end{equation}
\noindent i.e. as
\begin{equation}
R(\mathbf{q},i\omega_{l})=\int\frac{d^{3}\mathbf{k}}{(2\pi)^{3}}
\frac{n_{F}(\varepsilon_{\mathbf{k}})
-n_{F}(\varepsilon_{\mathbf{k+q}})}
{i\omega_{l}+\varepsilon_{\mathbf{k}}-\varepsilon_{\mathbf{k+q}}}
\:\left(|\mathbf{k}|\cos\theta+\frac{|\mathbf{q}|}{2}\right)^{2}
\:\:.
\end{equation}

By employing transformations analogous to
\eqref{expnf}-\eqref{tint} in the calculation of
$P^{0}_{\mathbf{q}}$ we arrive at the expression for
$R(\mathbf{q},i\omega_{l})$ in the zero-temperature limit :
\begin{equation}\label{}
R(\mathbf{q},i\omega_{l})\simeq \frac{k_{b}}{8(2\pi)^{2}v_{b}
|\mathbf{q}|^{3}} \int_{k_{b}+|\mathbf{q}|}^{k_{b}-|\mathbf{q}|}
\frac{t(t-k_{b})(t^{2}-k_{b}^{2})^{2}}
{t-k_{b}-i\frac{\omega_{l}}{v_{b}}}\:dt \:\:.
\end{equation}
\noindent By carrying out this integral and taking the static
limit $\omega_{l}\rightarrow 0$, we obtain
\begin{equation} 
R^{0}_{\mathbf{q}}\rightarrow -\frac{k_{b}^{4}}{6\pi^{2}v_{b}}
\qquad (\: |\mathbf{q}|\rightarrow 0 \:) \:\:.
\end{equation}

\bibliography{imbalance_aop}
\bibliographystyle{elsart-num}







\newpage

\newpage

\newpage

\newpage

\newpage

\newpage

\end{document}